\documentclass[pra,aps,twocolumn,showpacs,amsmath,amssymb,superscriptaddress]{revtex4}
\usepackage[dvips]{graphicx}
\usepackage{dcolumn}
\usepackage{bm}
\usepackage{subfigure}
\usepackage{setspace}
\usepackage{feynmp}
\usepackage{psfrag}
\usepackage{bbm}
\usepackage{ulem}
\usepackage{color}

\newcommand{\mbf}[1]{\boldsymbol{\mathbf{#1}}}

\newcommand{\be}{\begin{equation}}
\newcommand{\ee}{\end{equation}}
\newcommand{\ba}{\begin{array}}
\newcommand{\ea}{\end{array}}
\newcommand{\bqa}{\begin{eqnarray}}
\newcommand{\eqa}{\end{eqnarray}}

\newcommand{\bra}[1]{\ensuremath{\langle #1 |}}
\newcommand{\ket}[1]{\ensuremath{| #1 \rangle}}

\begin{document}

\title{Number-conserving master equation theory for a dilute Bose-Einstein condensate}
\author{Alexej Schelle}
\affiliation{Physikalisches Institut der
Albert-Ludwigs Universit\"{a}t Freiburg, 
Hermann-Herder-Str. 3, 
D-79104 Freiburg, 
Germany}
\affiliation{Laboratoire Kastler-Brossel, 
Universit\'e Pierre et Marie Curie-Paris 6, ENS, CNRS;  
4 Place Jussieu, 
F-75005 Paris, 
France}
\author{Thomas Wellens}
\affiliation{Physikalisches Institut der
Albert-Ludwigs Universit\"{a}t Freiburg, 
Hermann-Herder-Str. 3, 
D-79104 Freiburg, 
Germany}
\author{Dominique Delande}
\affiliation{Laboratoire Kastler-Brossel, 
Universit\'e Pierre et Marie Curie-Paris 6, ENS, CNRS;  
4 Place Jussieu, 
F-75005 Paris, 
France}
\author{Andreas Buchleitner}
\affiliation{Physikalisches Institut der
Albert-Ludwigs Universit\"{a}t Freiburg, 
Hermann-Herder-Str. 3, 
D-79104 Freiburg, 
Germany}
\date{\today}

\begin{abstract}
We describe the transition of $N$ weakly interacting 
atoms into a Bose-Einstein condensate within a 
number-conserving quantum master equation theory. 
Based on the separation of time scales for condensate  
formation and non-condensate thermalization, 
we derive a 
master equation 
for the condensate subsystem in the presence of 
the non-condensate environment under the inclusion of all two body interaction processes. 
We numerically monitor the condensate particle number distribution 
during condensate formation, and derive a condition under which the unique 
equilibrium steady state of a dilute, weakly interacting 
Bose-Einstein condensate is given 
by a Gibbs-Boltzmann thermal state of $N$ non-interacting atoms. 
\end{abstract}
\maketitle
\section{INTRODUCTION}	
\label{sectionI}
After almost one century of theoretical 
works to understand the existence, analysis and 
creation of a new state of matter at 
ultracold temperatures, the first experimental 
observation of a Bose-Einstein condensate was 
presented in Refs.~\cite{BEC_exp, BEC_exp1}. 
Nowadays, Bose-Einstein condensates are well established as 
a distinguished form of quantum matter enabling in situ studies 
of most disparate physical phenomena, such as Josephson oscillations~\cite{BEC_ObeJos}, or Anderson 
localization~\cite{BEC_Anderson, BEC_Anderson_2}, on a micrometer scale.

In the limit of zero temperature
and weak interactions,
where all atoms of the gas 
can be assumed to share 
the same single particle 
quantum state,  
the dynamics of the condensate is described accurately 
by the nonlinear Gross-Pitaevskii 
equation~\cite{Str_Pit_BEC}.
Finite temperature effects at thermal equilibrium are accounted for 
within higher order perturbation theories~\cite{Gri_Gapless, Castin4}. 
In contrast, only few theoretical works have been developed 
to model the non-equilibrium process of condensate formation itself.
Pioneering works, such as of Refs.~\cite{BEC_theo14, BEC_theo15, QKT, Walser1}, 
pointed primarily on the different dynamical stages of condensate formation
in terms of kinetic growth equations, 
and numerous efforts have led to highly 
accurate predictions for the time scale of condensate formation. 

Less is known about the 
condensate particle number \textit{distribution's dynamics} in 
a dilute, weakly interacting Bose gas consisting of a fixed number of $N$ particles. 
Another question under current study~\cite{Demler} is whether 
the equilibrium steady state of a dilute Bose-Einstein condensate --
which is 
finally reached only due to 
interatomic collisions, 
even in the 
case of very weak interactions --
is unique, and characterized by 
thermodymanics and
statistics of an ideal gas? 
And, how do quantum effects, such as 
number and energy fluctuations of the condensate and the non-condensate, which should become 
important especially for mesoscopic Bose-Einstein condensates, 
evolve during Bose-Einstein condensation 
and eventually drive the Bose gas 
into the final Gibbs-Boltzmann equilibrium state?
 
Here, we present a quantum master equation theory 
for a dilute Bose-Einstein condensate consisting of a fixed number 
of $N$ particles. 

In contrast to previously derived effective equations for the condensate dynamics under the influence of the non-condensate environment,
based for example on quantum kinetic theory \cite{QKT} or on an analogy with the laser master equation \cite{Kocharovsky},
our condensate master equation fully takes into account conservation of the total number of particles, i.e. of condensate plus non-condensate particles. In particular, the depletion of the non-condensate during the process of condensate formation results in condensate feeding and loss rates which are different from the case where the Bose gas is coupled to an external particle reservoir with a fixed chemical potential.

Apart from particle number conservation, our approach relies on the separation of time scales between the condensate and non-condensate dynamics. The time scale $\tau_0$ for condensate growth~\cite{formation, BEC_theo15_3, Schelle} is typically of the order of $1-4$ s, and thus much slower than the timescale  $\tau_{{\rm col}}\sim10-50~$ ms  of two-body collisions within the non-condensate \cite{Walser1, Griffin2, Hornberger1}. 
We assume that these collisions lead to an effective thermalization of the non-condensate within each subspace of fixed non-condensate particle number, and to a rapid decay of non-condensate correlation functions with a rate $\Gamma\simeq \tau_{{\rm col}}^{-1}$.
Under these conditions, particle exchange between condensate and non-condensate leads to a Markovian master equation for the
condensate's particle number distribution. Finally, we will show that its equilibrium steady state 
is  given by a Gibbs-Boltzmann thermal state of non-interacting 
particles under the condition that the Bose gas is sufficiently dilute, and that the decay of
non-condensate correlations does not occur too fast, i.e. $\hbar\beta\Gamma\ll 1$, 
where $\beta=1/k_{\mathrm{B}}T$ ($k_{\mathrm{B}}$ being the Boltzmann constant
and $T$ the temperature of the gas).

The paper is organized as follows: The derivation of the quantum master equation 
is given in section~\ref{chapter_micro}.
First, we define the condensate
and non-condensate Hamiltonians, and
decompose all two body 
interaction terms in a physically motivated way.
Using the assumptions mentioned above, we then 
derive the quantum master equation of Lindblad type 
for the reduced condensate density matrix. For dilute atomic gases in three-dimensional harmonic trapping potentials, 
the Lindblad master equation reduces to a simple rate equation for the condensate number distribution, 
describing transitions $N_0\to N_0\pm 1$ of the 
condensate particle number $N_0$.

In Sec.~\ref{sectionV}, the rate equation is used to study the dynamics of condensate formation, and its equilibrium steady state.
We numerically monitor the condensate particle 
number distribution during Bose-Einstein condensation and extract 
time scales for condensate formation.
The steady state solution of 
the rate 
equation finally yields a 
unique equilibrium steady state 
obeying detailed balance particle flow between condensate and 
non-condensate. 
In the 
case of weak interactions 
and under the condition $\hbar\beta\Gamma\ll 1$ of not too rapidly 
decaying non-condensate correlation functions, 
the steady state turns into a Gibbs-Boltzmann thermal state of a 
canonical ensemble of $N$ indistinguishable, 
non-interacting bosonic particles. 
 
We conclude in section~\ref{sectionVI}.

\section{Quantum master equation theory}
\label{chapter_micro}
The separation of time scales between non-condensate 
thermalization and condensate growth
motivates a decomposition of the gas into a 
condensate ``system'' and a non-condensate ``environment'' part,
see Sec.~\ref{section_subsystems}.
In Sec.~\ref{section_IA}, we then examine 
the two particle interactions between these 
subsystems as described by the Hamiltonian, Eq.~(\ref{Hamiltonian_full}). 
They fall into three different classes which we denote as
single particle, pair and scattering events.
 Under the inclusion of 
all these two body interaction processes,
the quantum master equation 
of Bose-Einstein condensation in a 
Bose gas 
with conserved particle number $N$ is finally derived in Secs.~\ref{section_evolution} and \ref{section_state}.

\subsection{Condensate and non-condensate subsystem}
\label{section_subsystems}
After defining the condensate mode, we examine the decomposition of the
full two body Hamiltonian in Eq.~(\ref{Hamiltonian_full}) into a condensate 
and a non-condensate part, and the interactions between them.
 
\subsubsection{Separation of the second-quantized field}
Quantitatively, we determine the condensate wave function $\Psi_{0}(\vec{\mbf{r}})$ 
(assuming all $N$ particles occupying the same condensate mode) 
by the Gross-Pitaevskii equation, 
\begin{equation}
\left[\frac{-\hbar^{2}\vec{\mbf{\nabla}}^{2}}{2m}+
V_{{\rm ext}}(\vec{\mbf{r}}) + 
gN\vert\Psi_{0}(\vec{\mbf{r}})\vert^{2}
-\mu_{0}\right]\Psi_{0}(\vec{\mbf{r}})=0\ ,
\label{timeindependent_GP}
\end{equation}
which, as discussed in Ref.~\cite{Castin4, Griffin1, Gardiner2}, gives a good approximation 
to the exact condensate mode at sufficiently low final temperatures, 
and sufficiently dilute atomic gases.

In our following treatment, 
we will use $\Psi_0(\vec{\mbf{r}})$ as defined by Eq.~(\ref{timeindependent_GP}) to describe
the condensate wave function also in a situation where initially not all particles occupy the condensate, and hence the condensate particle number will change as a function of time. Neglecting the associated time dependences of $\Psi_0(\vec{\mbf{r}})$ is justified
%tw in either one of the following cases: 
%tw In the limit of very weak interactions, since then 
%tw the condensate state is approximated, at all times, by the 
%tw ground state of the external trapping potential, $\ket{\chi_0}$, or, 
%tw if the initial state of the gas is already close to its equilibrium value.
%tw This is 
because it changes over the characteristic time scale for condensation,
much longer than the time scale $\tau_{{\rm col}}.$ We can thus 
%tw
employ
an adiabatic approximation and compute the
rates in the master equation for a fixed $\Psi_0$ and at the end
of the calculation only take into account their
dependence on $\Psi_0$, hence on time.  In the limit of very weak interactions, where 
the condensate state is approximated, at all times, by the 
ground state of the external trapping potential, $\ket{\chi_0}$, or, 
if the initial state of the gas is already close to its equilibrium value, the situation is even simpler as the $\Psi_0$ dependence can be entirely forgotten.

The total bosonic field $\hat{\Psi}$, expressed in an orthonormal basis $\lbrace\ket{\Psi_k}, k\in\mathbb{N}_0\rbrace$, where 
$\ket{\Psi_{0}}$ is the Gross-Pitaevskii ket, separates into
\begin{equation}
\hat{\Psi} = \vert\Psi_0\rangle\hat{a}_{0} + 
\sum_{k\ne0}\vert\Psi_k\rangle\hat{a}_{k}\ \equiv\hat{\Psi}_{0}
+ \hat{\Psi}_{\perp}\ ,
\label{field}
\end{equation}
with creation and annihilation 
operators $\hat{a}_{k}$ and $\hat{a}_{k}^{\dagger}$, respectively, 
satisfying usual bosonic commutation relations 
$\left[\hat{a}_k,\hat{a}^{\dagger}_l\right]=\delta_{kl}$, and 
$\left[\hat{a}_k,\hat{a}_l\right]=\left[\hat{a}^{\dagger}_k,\hat{a}^{\dagger}_l\right]=0$.

\subsubsection{Fock-Hilbert space}
The corresponding Fock states, forming a complete basis of the many particle Hilbert space $\mathcal F$ on which these 
operators act on are denoted 
by $\ket{N_{0},\lbrace N_{k}\rbrace}$.
The interpretation of a many particle 
Fock state is hence to find $N_{0}$ particles in the condensate 
mode $\ket{\Psi_0}$, and $\lbrace N_{k}\rbrace=\lbrace N_{1},N_{2},\ldots\rbrace$ particles 
in the modes $\lbrace\ket{\Psi_{1}},\ket{\Psi_{2}},\ldots\rbrace$.
The basis $\lbrace\ket{\Psi_k},k\in\mathbb{N}\rbrace$ is chosen such as to diagonalize the non-condensate Hamiltonian, see Eq.~(\ref{Hamiltonian_gas}).
Let us point out briefly the tensor structure of the total Fock-Hilbert space $\mathcal F$,
corresponding to the subsystems condensate and non-condensate, respectively:
\begin{equation}
\mathcal{F} = \mathcal{F}_{0}\otimes\mathcal{F}_{\perp}\ .
\label{Hilbert}
\end{equation}
As the condensate Hilbert space $\mathcal{F}_{0}$ is defined by $\mathcal{F}_{0} = 
{\rm span}\left\lbrace\vert N_{0}\rangle : N_{0}\in \mathbb{N}\right\rbrace$, so is the 
Hilbert space $\mathcal{F}_{\perp}$ of the non-condensate
by $\mathcal{F}_{\perp} = {\rm span}\lbrace\vert N_{1}, 
N_{2}, ...\rangle: N_{k}\in\mathbb{N}\rbrace$.
Partial traces will be taken according to Eq.~(\ref{Hilbert}) in the following.

\subsubsection{Decomposition of the Hamiltonian}
\label{section_deco}
The following decomposition of the Hamiltonian only requires
the validity of the Gross-Pitaevskii equation for the condensate mode, and 
the orthogonality of the two fields $\hat{\Psi}^{\dagger}_{0}$ and 
$\hat{\Psi}_{\perp}$, in the sense that
\begin{equation}
\int{\rm d}\vec{\mbf{r}}~\hat{\Psi}^{\dagger}_{0}(\vec{\mbf{r}})\hat{\Psi}_{\perp}(\vec{\mbf{r}}) = 0\ .
\label{orthogonal}
\end{equation}
The Hamiltonian $\hat{\mathcal{H}}$ in 
second quantization, including two body interactions~\cite{Str_Pit_BEC}, is given by 
\begin{equation}
\begin{split}
\hat{\mathcal{H}} = &\int\,{\rm d}\vec{\mbf{r}}~\hat{\Psi}^{\dagger}(\vec{\mbf{r}})
\left[-\frac{\hbar^{2}\vec{\mbf{\nabla}}^{2}}{2m}+V_{{\rm ext}}(\vec{\mbf{r}})\right]
\hat{\Psi}(\vec{\mbf{r}}) \\ 
+ &\frac{g}{2}\int\,{\rm d}\vec{\mbf{r}}~
\hat{\Psi}^{\dagger}(\vec{\mbf{r}})\hat{\Psi}^{\dagger}(\vec{\mbf{r}})
\hat{\Psi}(\vec{\mbf{r}})
\hat{\Psi}(\vec{\mbf{r}})\ ,
\label{Hamiltonian_full}
\end{split}
\end{equation}
where $\hat{\Psi}(\vec{\mbf{r}})= \hat{\Psi}_0(\vec{\mbf{r}}) + 
\hat{\Psi}_\perp(\vec{\mbf{r}})$ denotes the second-quantized bosonic field, and 
with $g=4\pi a\hbar^2/m$ quantifying the interaction strength in terms of the 
s-wave scattering length $a$.
The neglect of three-body collisions implied by Eq.~(\ref{Hamiltonian_full}) is justified in the dilute regime $a\varrho^{1/3}\ll 1$, with $\varrho$ the density of the atomic gas.
The field decomposition in Eq.~(\ref{field}) splits the Hamiltonian $\hat{\mathcal{H}}$ into 
three basic contributions, 
\begin{equation}
\hat{\mathcal{H}} =  \hat{\mathcal{H}}_{0} + \hat{\mathcal{H}}_{\perp} + \hat{\mathcal{V}}_{0\perp}\ , 
\label{Hamiltonian}
\end{equation}
where $\hat{\mathcal{H}}_{0}$ and $\hat{\mathcal{H}}_{\perp}$ 
describe a pure condensate and non-condensate, respectively. 

The condensate Hamiltonian $\hat{\mathcal{H}}_{0}$ contains the single particle 
contribution linear in the field $\hat{\Psi}_0$, 
as well as the nonlinear, self-interacting two body term, 
and is given by
\begin{eqnarray}
\hat{\mathcal{H}}_{0} &  = &\int{\rm d}\vec{\mbf{r}}~\hat{\Psi}_{0}^{\dagger}(\vec{\mbf{r}})
\left[-\frac{\hbar^{2}\vec{\mbf{\nabla}}^{2}}{2m}+V_{{\rm ext}}(\vec{\mbf{r}})\right]
\hat{\Psi}_{0}(\vec{\mbf{r}}) \nonumber \\
& & + \frac{g}{2}\int{\rm d}\vec{\mbf{r}}~\hat{\Psi}_{0}^{\dagger}(\vec{\mbf{r}})
\hat{\Psi}_{0}^{\dagger}(\vec{\mbf{r}})
\hat{\Psi}_{0}(\vec{\mbf{r}})\hat{\Psi}_{0}(\vec{\mbf{r}})\ .
\label{Hamiltonian_condensate}
\end{eqnarray}
When the average number of particles in the condensate
is much larger than unity, a mean field approximation can be used to compute
the ground state of $\hat{\mathcal{H}}_{0}$, which
allows to recover the ordinary Gross-Pitaevskii equation (\ref{timeindependent_GP}).

Concerning the Hamiltonian of the background gas, 
$\hat{\mathcal{H}}_{\perp}$, we first write down 
the contribution bilinear in the non-condensate fields $\hat{\Psi}_\perp$ 
and $\hat{\Psi}_\perp^\dagger$, respectively:
\begin{eqnarray}
\hat{\mathcal{H}}_{\perp}  & = &\int{\rm d}\vec{\mbf{r}}~\hat{\Psi}_{\perp}^{\dagger}(\vec{\mbf{r}})
\left[-\frac{\hbar^{2}\vec{\mbf{\nabla}}^{2}}{2m}+V_{{\rm ext}}(\vec{\mbf{r}})\right]
\hat{\Psi}_{\perp}(\vec{\mbf{r}})\nonumber \\
& & = \sum_{k\ne0}\epsilon_k\hat{a}^{\dagger}_k\hat{a}_k \ ,
\label{Hamiltonian_gas}
\end{eqnarray}
where $\epsilon_k$ are single particle energies of non-condensate particles.
To model interactions between non-condensate particles, we assume that these lead to a rapid 
thermalization in the non-condensate thermal vapor,
as will be further discussed in Sec.~\ref{section_rapid} below.

Finally, the last term in Eq.~(\ref{Hamiltonian}), $\hat{\mathcal{V}}_{0\perp}$, 
describes all two body interactions between condensate and non-condensate. 
This term will be examined in the following subsection. 

\subsection{Two-body interaction processes}
\label{section_IA}
Inserting the decomposition of the field $\hat{\Psi}(\vec{\mbf{r}})$, 
Eq.~(\ref{field}), into the Hamiltionian $\hat{\mathcal{H}}$, Eq.~(\ref{Hamiltonian_full}), we find, 
besides the condensate and non-condensate 
Hamiltonians, Eqs.~(\ref{Hamiltonian_condensate},~\ref{Hamiltonian_gas}), various terms describing 
two particle interaction processes. 
Sorting these according to the number of condensate and non-condensate particles, which
are created or annihilated during a two body collision event, we obtain
\begin{equation}
\hat{\mathcal{V}}_{\perp0} = \hat{\mathcal{V}}_{\rightsquigarrow} + 
\hat{\mathcal{V}}_{\leftrightsquigarrow} + \hat{\mathcal{V}}_{\circlearrowright}\ ,
\label{Hamiltonian_interaction}
\end{equation}
where
\begin{equation}
\begin{split}
 \hat{\mathcal{V}}_{\rightsquigarrow} = ~&g\int{\rm d}\vec{\mbf{r}}~
\hat{\Psi}_{\perp}^{\dagger}(\vec{\mbf{r}})
\hat{\Psi}_{\perp}^{\dagger}(\vec{\mbf{r}})\hat{\Psi}_{\perp}(\vec{\mbf{r}})
\hat{\Psi}_{0}(\vec{\mbf{r}}) \\
+~&g\int{\rm d}\vec{\mbf{r}}~\hat{\Psi}_{0}^{\dagger}(\vec{\mbf{r}})
\hat{\Psi}_{\perp}^{\dagger}(\vec{\mbf{r}})\hat{\Psi}_{\perp}(\vec{\mbf{r}})
\hat{\Psi}_{\perp}(\vec{\mbf{r}})\  
\label{IA}
\end{split}
\end{equation}
accounts for single particle events, where the condensate particle 
number changes by $\Delta N_0=\pm 1$, and correspondingly, the
number of non-condensate particles by $\Delta N_\perp=\mp 1$.

Second,
\begin{equation}
\begin{split}
\hat{\mathcal{V}}_{\leftrightsquigarrow} = ~&\frac{g}{2}\int\,{\rm d}\vec{\mbf{r}}~
\hat{\Psi}_{\perp}^{\dagger}(\vec{\mbf{r}})
\hat{\Psi}_{\perp}^{\dagger}(\vec{\mbf{r}})\hat{\Psi}_{0}(\vec{\mbf{r}})
\hat{\Psi}_{0}(\vec{\mbf{r}}) \\
+~&\frac{g}{2}\int\,{\rm d}\vec{\mbf{r}}~\hat{\Psi}_{0}^{\dagger}(\vec{\mbf{r}})
\hat{\Psi}_{0}^{\dagger}(\vec{\mbf{r}})\hat{\Psi}_{\perp}(\vec{\mbf{r}})
\hat{\Psi}_{\perp}(\vec{\mbf{r}})\  
\label{IAA}
\end{split}
\end{equation}
describes pair events, where two condensate particles are created or annililated, i.e., $\Delta N_0=\pm 2$ and $\Delta N_\perp=\mp 2$.

Finally, the term
\begin{equation}
\hat{V}_{\circlearrowright} = 2g\int{\rm d}\vec{\mbf{r}}{\rm~}
\hat{\Psi}_{\perp}^{\dagger}(\vec{\mbf{r}})\hat{\Psi}^{\dagger}_{0}(\vec{\mbf{r}})
\hat{\Psi}_{\perp}(\vec{\mbf{r}})\hat{\Psi}_{0}(\vec{\mbf{r}})\ 
\label{IAAA}
\end{equation}
describes scattering events, where the number of condensate and non-condensate particles is unchanged
($\Delta N_0=\Delta N_\perp=0$). As we will see later, scattering events do not contribute to the master equation for the condensate density matrix,
which will mainly be governed by single particle events, with negligible influence of pair events.
\begin{figure}[t]
\vspace*{1.0cm}
\begin{center}
\begin{fmffile}{condensate_induced}
\begin{fmfchar*}(100,50)
   \fmfleft{hm,hp}
   \fmf{fermion}{hm,Zhh,hp}
   \fmflabel{$0$}{hm}
   \fmflabel{$k^{\star}$}{hp}
   \fmf{boson,label=$\rightsquigarrow$}{Zhh,Zkk}
   \fmf{fermion}{kb,Zkk,k}
   \fmfright{kb,k}      
   \fmflabel{$0$}{kb}
   \fmflabel{$0^{\star}$}{k}
   \fmfdot{Zhh,Zkk}
 \end{fmfchar*}
\end{fmffile}
\nolinebreak
\hspace{0.5cm}
\begin{fmffile}{environment-induced}
\begin{fmfchar*}(100,50)
   \fmfleft{hm,hp}
   \fmf{fermion}{hm,Zhh,hp}
   \fmflabel{$0$}{hm}
   \fmflabel{$k^{\star}$}{hp}
   \fmf{boson,label=$\rightsquigarrow$}{Zhh,Zkk}
   \fmf{fermion}{kb,Zkk,k}
   \fmfright{kb,k}      
   \fmflabel{$m$}{kb}
   \fmflabel{$l^{\star}$}{k}
   \fmfdot{Zhh,Zkk}
 \end{fmfchar*}
\end{fmffile}
\end{center}
\vspace*{1.0cm}
\begin{center}
\begin{fmffile}{pair_events}
\begin{fmfchar*}(100,50)
   \fmfleft{hm,hp}
   \fmf{fermion}{hm,Zhh,hp}
   \fmflabel{$0$}{hm}
   \fmflabel{$k^{\star}$}{hp}
   \fmf{boson,label=$\leftrightsquigarrow$}{Zhh,Zkk}
   \fmf{fermion}{kb,Zkk,k}
   \fmfright{kb,k}      
   \fmflabel{$0$}{kb}
   \fmflabel{$l^{\star}$}{k}
   \fmfdot{Zhh,Zkk}
 \end{fmfchar*}
\end{fmffile}
\nolinebreak
\hspace{0.5cm}
\begin{fmffile}{scattering}
 \begin{fmfchar*}(100,50)
   \fmfleft{em,ep}
   \fmf{fermion}{em,Zee,ep}
   \fmflabel{$l$}{em}
   \fmflabel{$0^{\star}$}{ep}
   \fmf{boson,label=$\circlearrowright$}{Zee,Zff}
   \fmf{fermion}{fb,Zff,f}
   \fmfright{fb,f}      
   \fmflabel{$0$}{fb}
   \fmflabel{$k^{\star}$}{f}
   \fmfdot{Zee,Zff}
\end{fmfchar*}
\end{fmffile}
\end{center}
\vspace{1.0cm}
\caption{Diagrammatic representation of all
two body loss processes in Eq.~(\ref{Hamiltonian_interaction}). 
The upper two diagrams show single particle 
losses ($\rightsquigarrow$), where one non-condensed particle 
is created and one condensate atom is annihilated. The upper left diagram denotes a
term linear in the non-condensed 
field, $\mathcal{O}(\hat{\Psi}_{\perp})$, which vanishes
in combination with crossed 
single particle terms as a consequence 
of the Gross-Pitaevskii Eq.~(\ref{timeindependent_GP}) and the orthogonality condition
in Eq.~(\ref{orthogonal}). 
The lower diagrams show pair losses ($\leftrightsquigarrow$, lower left) 
and scattering processes ($\circlearrowright$, lower right). 
The conjugate processes (not shown),
related to condensate feeding, are obtained 
by exchanging the corresponding labels 
with respect to the diagram center.}
\label{diagrams}
\end{figure}
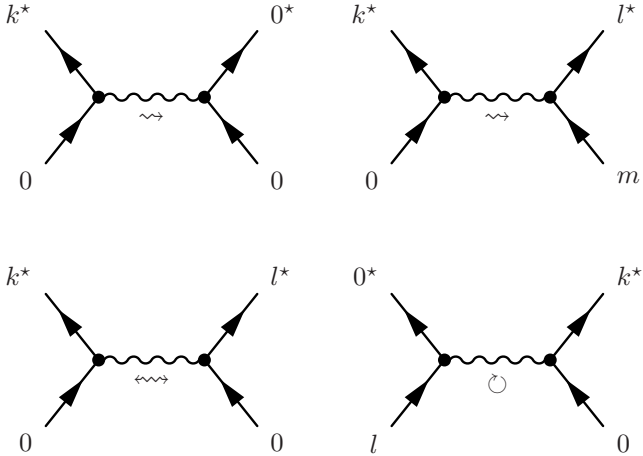

To illustrate the different interaction terms, we introduce a diagrammatic representation of 
the interaction matrix elements \cite{Castin5,Tan_Dup_Gry}. These are depicted in Fig.~\ref{diagrams}: 
Annihilation and creation of condensate particles
are denoted by $0$ and $0^*$, respectively, whereas $k,l,m$ and $k^*,l^*,m^*$ refer to annihilated, 
or created particles of the corresponding non-condensate modes. Note that 
Fig.~\ref{diagrams} contains only condensate loss events, where the number of condensate particles decreases. The conjugate processes,
corresponding to condensate feeding, are obtained 
by exchanging the corresponding labels 
with respect to the diagram center. 

Furthermore, Fig.~\ref{diagrams} also shows processes of first order in the non-condensate 
field (upper left diagram), which are, however, not contained in Eq.~(\ref{Hamiltonian_interaction}). The reason is that these processes
cancel 
out with mixed, single particle contributions between condensate and non-condensate fields 
in the Hamiltonian in Eq.~(\ref{Hamiltonian_full}). 
This is a consequence of the orthogonality of the two fields, 
$\hat{\Psi}^{\dagger}_{0}(\vec{\mbf{r}})$ and $\hat{\Psi}_{\perp}(\vec{\mbf{r}})$, see~Eq.~(\ref{orthogonal}), and 
the fact that $\Psi_0(\vec{\mbf{r}})$ is an approximate solution 
of the Gross-Pitaevskii equation, Eq.~(\ref{timeindependent_GP}), 
for sufficiently low temperatures (i.e., a sufficiently peaked 
condensate number distribution close to $N$). 
Indeed, when we combine upper left diagrams in Fig.~\ref{diagrams}, 
and their hermitian conjugates, with mixed single particle 
contributions in Eq.~(\ref{Hamiltonian_full}), we get the vanishing term 
\begin{eqnarray}
& & \int{\rm d}\vec{\mbf{r}}~\hat{\Psi}^{\dagger}_\perp(\vec{\mbf{r}})\left[
\frac{-\hbar^2\vec{\mbf{\nabla}}^2}{2m} + V_{{\rm ext}}(\vec{\mbf{r}}) + 
\hat{\Psi}^{\dagger}_0(\vec{\mbf{r}})\hat{\Psi}_0(\vec{\mbf{r}})\right]
\hat{\Psi}_0(\vec{\mbf{r}})  \nonumber\\
& & \simeq~\sum_{k\neq 0} \hat{a}^\dagger_k
\int{\rm d}\vec{\mbf{r}}~\Psi_k^*(\vec{\mbf{r}}) \Psi_0(\vec{\mbf{r}}) \mu_0 \hat{a}_0  = 0\ . 
\end{eqnarray}
In total, the Hamiltonian $\hat{\mathcal{H}}$ in Eq.~(\ref{Hamiltonian_full}) thus decomposes into 
\begin{equation}
\hat{\mathcal{H}} = \hat{\mathcal{H}}_{0} + \hat{\mathcal{H}}_{\perp}
+ \hat{\mathcal{V}}_{\rightsquigarrow} + 
\hat{\mathcal{V}}_{\leftrightsquigarrow} + 
\hat{\mathcal{V}}_{\circlearrowright}\ ,
\label{Hamiltonian_decomposition}
\end{equation}
where the different interaction terms $\hat{\mathcal{V}}_{\rightsquigarrow}$ in Eq.~(\ref{IA}), 
$\hat{\mathcal{V}}_{\leftrightsquigarrow}$ in Eq.~(\ref{IAA}), and $\hat{\mathcal{V}}_{\circlearrowright}$ in Eq.~(\ref{IAAA}), 
account for single particle ($\rightsquigarrow$), pair ($\leftrightsquigarrow$) and 
scattering ($\circlearrowright$) contributions.

\subsection{Evolution equation for the total density matrix}
\label{section_evolution}
In analogy to the standard quantum optical 
derivation \cite{Tan_Dup_Gry, Gardiner}, we start with 
the von-Neumann equation, considering a 
many particle state $\hat{\sigma}^{(N)}(t)$
of fixed particle number $N$, 
defined on the Fock-Hilbert space 
$\mathcal{F}=\mathcal{F}_0\otimes\mathcal{F}_\perp$
in Eq.~(\ref{Hilbert}):
\begin{equation}
\frac{\partial \hat{\sigma}^{(N)}(t)}{\partial t}  = 
-\frac{i}{\hbar}\left[\hat{\mathcal{H}},\hat{\sigma}^{(N)}(t)\right]\ ,
\label{Liouville}
\end{equation}
where $\hat{\mathcal{H}}$ is the total Hamiltonian, see Eq.~(\ref{Hamiltonian_full}). 
With the decomposition of $\hat{\mathcal{H}}$ in Eq.~(\ref{Hamiltonian}), the von-Neumann equation turns into
\begin{equation}
\begin{split}
\frac{\partial\hat{\sigma}^{(N)}(t)}{\partial t} = 
&-\frac{i}{\hbar}\left[\hat{\mathcal{H}}_{0},\hat{\sigma}^{(N)}(t)\right]-
\frac{i}{\hbar}\left[\hat{\mathcal{H}}_{\perp},\hat{\sigma}^{(N)}(t)\right] \\
&-\frac{i}{\hbar}\left[\hat{\mathcal{V}}_{0\perp},\hat{\sigma}^{(N)}(t)\right]\ . \\
\end{split}
\end{equation}
Note that we use here the linearized non-condensate Hamiltonian in Eq.~(\ref{Hamiltonian_gas}). 
We transform all operators, i.e., the condensate 
and the non-condensate field, 
$\hat{\Psi}_0(\vec{\mbf{r}})$ and $\hat{\Psi}_\perp(\vec{\mbf{r}})$, as well 
as the density matrix $\hat{\sigma}^{(N)}(t)$, 
to the interaction picture (denoted by the label $I$), 
which is carried out 
with respect to the Hamiltonian parts $\hat{\mathcal{H}}_{0}$ and $\hat{\mathcal{H}}_{\perp}$ 
of the subsystems condensate and non-condensate. 
The different operators hence undergo the transformation
\begin{equation}
 \hat{X}(t) \rightarrow \hat{X}^{(I)}(t) = \hat{\mathcal{U}}(t)\hat{X}\hat{\mathcal{U}}^{\dagger}(t)\ ,
\end{equation}
with respect to the time evolution operator $\hat{\mathcal{U}}(t)$ given by
\begin{equation}
\hat{\mathcal{U}}(t) = {\rm exp}\left[\frac{i}{\hbar}
\left(\hat{\mathcal{H}}_{0}+\hat{\mathcal{H}}_{\perp}\right)t\right]\ .
\label{trafo_int}
\end{equation}
The time evolution of the full density operator 
$\hat{\sigma}^{(N,I)}(t)$ in the interaction picture is then determined by the 
interaction between condensate and non-condensate particles, according to:
\begin{equation}
\frac{\partial\hat{\sigma}^{(N,I)}(t)}{\partial t}  = 
-\frac{i}{\hbar}[\hat{\mathcal{V}}^{(I)}_{0\perp}(t),\hat{\sigma}^{(N,I)}(t)]\ .
\label{int}
\end{equation}
where $\hat{\mathcal{V}}^{(I)}_{0\perp}(t)$ is obtained by inserting the 
time dependent annihilation (and creation) operators, e.g. $a_0(t)=a_0 e^{-i\mu_0 t/\hbar}$ and
$a_k(t)=a_k e^{-i\epsilon_k t/\hbar}$, in the corresponding time 
independent expressions derived in the previous section. 
Integration of Eq.~(\ref{int}) between $t$ and $t+\Delta t$ leads to 
\begin{equation}
\begin{split}
\hat{\sigma}^{(N,I)}(t+\Delta t) &= \hat{\sigma}^{(N,I)}(t) \\ 
-&\frac{i}{\hbar}\int\limits_{t}^{t+\Delta t}{\rm d}t^{\prime}
\left[\hat{\mathcal{V}}_{0\perp}^{(I)}(t^{\prime}),\hat{\sigma}^{(N,I)}(t^{\prime})\right]\ . 
\end{split}
\label{integral}
\end{equation}
For short times, $\Delta t$, a good approximate solution of Eq.~(\ref{integral}) is obtained by its iteration
up to second order in $\hat{\mathcal{V}}_{0\perp}$ 
(which is required since the first-order terms vanish, as we will see later): 
\vspace{-0.4cm}
\begin{equation}
\begin{split}
& \Delta\hat{\sigma}^{(N,I)}(t) = -\frac{i}{\hbar}\int\limits_{t}^{t+\Delta t}{\rm d}t^{\prime}\left[\hat{\mathcal{V}}^{(I)}_{0\perp}(t^{\prime})
,\hat{\sigma}^{(N,I)}(t)\right] \\ 
&-\int\limits_{t}^{t+\Delta t}
{\rm d}t^{\prime}\int\limits_{t}^{t^{\prime}}\frac{{\rm d}t^{\prime\prime}}
 {\hbar^{2}}\left[\hat{\mathcal{V}}^{(I)}_{0\perp}(t^{\prime}),
 \left[\hat{\mathcal{V}}^{(I)}_{0\perp}(t^{\prime\prime}),
\hat{\sigma}^{(N,I)}(t)\right]\right]\ , 
\label{non_split_Master}
\end{split}
\end{equation}
where we have set $\Delta\hat{\sigma}^{(N,I)}(t) = \hat{\sigma}^{(N,I)}(t+\Delta t)-\hat{\sigma}^{(N,I)}(t)$.
Note that Eq.~(\ref{non_split_Master}) expresses 
the state at time $t+\Delta t$ (left-hand side) fully, 
as a function of the state $\hat{\sigma}^{(N,I)}(t)$ at time $t$ - 
in contrast to the exact Eq.~(\ref{integral}), 
where states $\hat{\sigma}^{(N,I)}(t')$ at all intermediate times $t'$ appear on the right-hand side.

\subsection{Time evolution of the reduced condensate density matrix}
\label{section_state}
The time evolution of the condensate in the presence of the non-condensate gas
is obtained by taking the partial trace over $\mathcal{F}_\perp$ in Eq.~(\ref{non_split_Master}).
To get a Markovian master equation for the reduced condensate 
density matrix, $\hat{\rho}^{(N)}_0(t)={\rm Tr}_{\mathcal{F}_\perp}\hat{\sigma}^{(N)}(t)$,
we use a Born-Markov ansatz generalized for the $N$-particle state $\sigma^{(N)}(t)$ which allows to express
$\sigma^{(N)}(t)$ completely in terms of the reduced 
condensate density matrix $\hat{\rho}^{(N)}_0(t)$ at time $t$, see Eq.~(\ref{Bose_state_final}).

\subsubsection{Non-condensate thermalization}
\label{section_rapid}
In standard derivations of master equations for systems coupled
 to thermal reservoirs~\cite{Tan_Dup_Gry, QKT}, the Markov assumption 
 is justified by assuming a thermal
state $\hat{\rho}_E(T)$ for the environment, which is supposed to be 
unchanged by the interaction with the system. Then, the 
total state $\hat{\sigma}(t)$ would be given by the product 
$\hat{\sigma}(t) = \hat{\rho}_0(t)\otimes\hat{\rho}_{E}(T)$, 
hence completely determined by the reduced state $\hat{\rho}_0(t)$ of the (condensate) 
subsystem. 
However, in our case, this 
simple product ansatz cannot be applied, since condensate and 
non-condensate are correlated by particle number conservation:
If one finds $N_0$ particles in the condensate, the particle number in the non-condensate is 
determined as $N_\perp=N-N_0$, and vice versa.

\begin{figure}[t]
\begin{center}
\includegraphics[width=7.2cm,height=4.8cm,angle = 0.0]{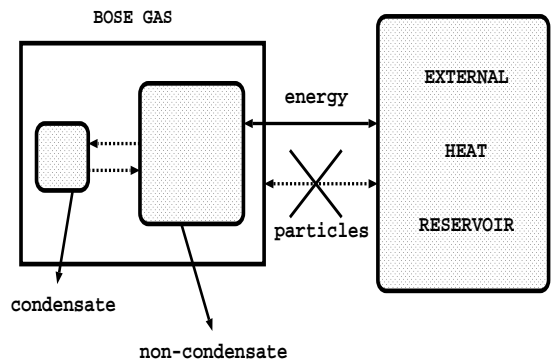}
\caption{Schematics of microsopic many particle dynamics. 
The total number of atoms in the Bose gas is fixed to $N$ and conserved during condensate formation. 
Atomic collisions \textit{within} the non-condensate 
are modelled by coupling the non-condensate part of the gas to a heat reservoir which is at fixed temperature $T$. 
The condensate part is initially not at equilibrium with the non-condensed fraction, 
and both systems undergo a net exchange of particles, induced by atomic two body collisions 
\textit{between} condensate and non-condensate atoms, which are fully taken into account 
in the derivation of the master equation. 
The final equilibrium steady state of the gas 
exhibits detailed balance particle flow between 
condensate and non-condensate.}
\label{topology}
\end{center}
\end{figure} 
The physical origin of the non-condensate thermalization is the interaction between 
non-condensate particles, which leaves the number of non-condensate particles unchanged. 
We hence couple the non-condensate to a heat 
bath only allowing for exchange of energy, but not of particles, see Fig.~\ref{topology}. The thermalization then occurs only within subspaces 
of fixed particle number. In addition, we assume that coherences between subspaces of different particle number 
are destroyed due to the coupling with the heat bath. Under this assumption, the total $N$-particle state is obtained as: 
\begin{equation}
\hat{\sigma}^{(N)}(t) = \sum_{N_{0}=0}^{N}p_N(N_0,t)\ket{N_0}\bra{N_0}
\otimes\hat{\rho}_{\perp}(N-N_{0},T)\ ,
\label{Bose_state_final}
\end{equation}
where $p_N(N_0,t)$ denotes the probability of finding $N_0$ particles in the condensate (or, equivalently, $N-N_0$ particles in the non-condensate), and 
\begin{equation}
\hat{\rho}_{\perp}(N-N_{0},T) = \frac{\hat{\mathcal{Q}}_{N-N_{0}}
{\rm e}^{-\beta\hat{\mathcal{H}}_{\perp}}\hat{\mathcal{Q}}_{N-N_{0}}}{\mathcal{Z}_\perp(N-N_{0},T)}\ 
\label{NC_thermal}
\end{equation}
describes a thermal state projected onto the subspace of $N-N_0$ non-condensate particles, with corresponding projector
$\hat{\mathcal{Q}}_{N-N_{0}}$, and normalization factor~\cite{Statistics}
\begin{equation}
\mathcal{Z}_\perp(N-N_{0},T) = {\rm Tr}_{\mathcal{F}_\perp}\left\lbrace
\hat{\mathcal{Q}}_{N-N_{0}}{\rm e}^{-\beta\hat{\mathcal{H}}_{\perp}}
\hat{\mathcal{Q}}_{N-N_{0}}\right\rbrace\ .
\label{partition}
\end{equation}
Note that, since $\hat{\sigma}^{(N)}(t)$ is diagonal in the Fock basis, it is invariant 
under the free evolution $\hat{\mathcal{U}}(t)$, Eq.~(\ref{trafo_int}), 
and hence $\hat{\sigma}^{(N,I)}(t)= \hat{\sigma}^{(N)}(t)$. In the following, 
we hence drop the index \lq $I$\rq\ referring to the interaction picture for the $N$-particle state $\hat{\sigma}^{(N)}(t)$, 
or its reduced condensate state $\hat{\rho}^{(N)}_{0}(t)$, see below.

\subsubsection{Evolution equation for the condensate density matrix}
Taking the partial trace over the non-condensate, we obtain 
the reduced condensate density matrix: 
\begin{equation}
\hat{\rho}^{(N)}_{0}(t) = {\rm Tr}_{\mathcal{F}_\perp}
\left\lbrace\hat{\sigma}^{(N)}(t)\right\rbrace = 
\sum_{N_{0}=0}^{N}p_N(N_{0},t)\ket{N_0}\bra{N_0}\ .
\label{state_C}
\end{equation}
Obviously, also the reduced condensate state 
is diagonal in particle number representation 
as a direct consequence of our assumptions on particle 
number conservation and rapid non-condensate thermalization. 
Thus, both, the reduced condensate density matrix, Eq.~(\ref{state_C}), as well as the total $N$-particle state,
Eq.~(\ref{Bose_state_final}), are completely determined by the 
condensate particle number distribution $p_N(N_0,t)$.

Inserting Eq.~(\ref{Bose_state_final}) in the right-hand side of Eq.~(\ref{non_split_Master}), 
and taking the partial trace over the
non-condensate, leads to a closed 
evolution equation for the reduced condensate density matrix. 
Moreover, it can be shown that the terms of first order in the interaction $\hat{\mathcal V}_{0\perp}^{(I)}$ 
vanish after taking the partial trace over $\mathcal{F}_\perp$.
Indeed, from the diagonal form of the $N$-particle state $\hat{\sigma}^{(N)}(t)$, see Eq.~(\ref{Bose_state_final}), it follows that
\begin{equation}
{\rm Tr}_{\mathcal{F}_\perp}\left\lbrace
\left[\hat{\mathcal{V}}^{(I)}_{0\perp}(t^{\prime})
,\hat{\sigma}^{(N,I)}(t)\right]
\right\rbrace=0\ .
\end{equation}
From the remaining second order terms, we obtain: 
\begin{widetext}
\begin{equation}
  \Delta\hat{\rho}^{(N)}_{0}(t) = -\sum_{N_{0}=0}^{N}
 \int\limits_{t}^{t+\Delta t}{\rm d}t^{\prime}\int\limits_{t}^{t^{\prime}}\frac{{\rm d}
 t^{\prime\prime}}{\hbar^{2}}{\rm~Tr}_{\mathcal{F}_\perp}\left[\hat{\mathcal{V}}^{(I)}_{0\perp}(t^{\prime}) ,
  \left[\hat{\mathcal{V}}^{(I)}_{0\perp}(t^{\prime\prime}),
  p_N(N_{0},t)\ket{N_0}\bra{N_0}\otimes\hat{\rho}_{\perp}(N-N_{0},T)\right]\right]\ .\\
\label{Mast}
\end{equation}
\end{widetext}
Writing the interaction term $\hat{\mathcal{V}}^{(I)}_{0\perp}(t^{\prime})$ 
as a sum over the three different processes (single particle, pair and scattering events), 
see Eq.~(\ref{Hamiltonian_interaction}), we can now verify that any mixed commutator 
in Eq.~(\ref{Mast}) vanishes - again as a consequence of the diagonality of $\hat{\sigma}^{(N)}(t)$. 

Hence,
single particle, pair and scattering events in the gas are
dynamically independent from each other. Furthermore, it can be shown 
that scattering events, described by $\hat{\mathcal{V}}_{\circlearrowright}$, do not contribute, 
since they leave the number of condensate particles unchanged. We are left with: 
\begin{equation}
  \frac{\Delta\hat{\rho}^{(N)}_{0}(t)}{\Delta t} 
= \left.\frac{\Delta\hat{\rho}^{(N)}_{0}(t)}{\Delta t}\right\vert_{\rightsquigarrow}
  +\left.\frac{\Delta\hat{\rho}^{(N)}_{0}(t)}{\Delta t}\right\vert_{\leftrightsquigarrow}\ ,
\label{Mast_2}
\end{equation}
where the two terms on the right-hand side of Eq.~(\ref{Mast_2}) are obtained by inserting the corresponding interaction terms 
$\hat{\mathcal{V}}_{\rightsquigarrow}$ and $\hat{\mathcal{V}}_{\leftrightsquigarrow}$,
instead of the full interaction $\hat{\mathcal{V}}_{\perp0}$ into Eq.~(\ref{Mast}). 

\subsubsection{Quantum master equation of Lindblad type}
\label{section_master}
In order to perform the time integration in Eq.~(\ref{Mast}), we first notice that 
the right-hand side depends only on the time difference $\tau=t'-t''$.
Second, we assume that only times $\tau\ll \Delta t$ contribute to the integral, 
due to the rapid decay of non-condenate correlation functions.
To implement this rapid decay, we assume that the two point correlation functions 
of the non-condensate decay on the 
average time scale $\tau_{{\rm col}}$ of a two body collision event. 
Performing the time integral as
\begin{widetext}
\begin{equation}
\int_t^{t+\Delta t}{\rm d}t \int_t^{t'} {\rm d}t''=\int_0^{\Delta t}{\rm d}\tau \int_{t+\tau}^{t+\Delta t}{\rm d}t'\simeq \Delta t\int_0^{\Delta t}{\rm d}\tau\ ,
\end{equation}
\end{widetext}
using $\tau\ll \Delta t$ for the second equality. However, even though $\Delta t$ 
has to be larger than the decay time $\Gamma^{-1}\simeq \tau_{{\rm col}}$ of non-condensate correlation functions (see below), 
it is still much smaller than the time scale $\tau_0$ for the condensate evolution.
In this case, the coarse-grained rate $\Delta\hat{\rho}_0(t)/\Delta t$ can be replaced by the 
instantaneous time derivative $\partial\hat{\rho}_0(t)/\partial t$ to 
obtain the following Lindblad master equation: 
\begin{widetext}
\begin{equation}
\begin{split}
\frac{\partial\hat{\rho}^{(N)}_{0}(t)}{\partial t} &=\sum^{N}_{N_{0}=0,\atop{j=+,-}} 
\xi_N^j(N_0,T)\left[\hat{\mathcal{S}}_j(N_0)\hat{\rho}^{(N)}_{0}(t)\hat{\mathcal{S}}_j^\dagger(N_0)
-\frac{1}{2}\left\lbrace\hat{\mathcal{S}}_j^\dagger(N_0)\hat{\mathcal{S}}_j(N_0),
\hat{\rho}^{(N)}_{0}(t)\right\rbrace_{+}\right]\\ 
&+\sum^{N}_{N_{0}=0,\atop{j=+,-}}\gamma_N^j(N_0,T)\left[\hat{\mathcal{P}}_j(N_0)\hat{\rho}^{(N)}_{0}(t)\hat{\mathcal{P}}_j^{\dagger}(N_0)
-\frac{1}{2}\left\lbrace\hat{\mathcal{P}}_j^{\dagger}(N_0)\hat{\mathcal{P}}_j(N_0),
\hat{\rho}^{(N)}_{0}(t)\right\rbrace_{+}\right] \ ,
\end{split}
\label{Lindblad_final}
\end{equation}
where the quantum jump operators $\hat{\mathcal{S}}_{\pm}(N_0)$, 
and $\hat{\mathcal{P}}_{\pm}(N_0)$
are defined by 
\begin{equation}
\begin{split}
&\hat{\mathcal{S}}_+(N_0)=\ket{N_0+1}\bra{N_0},{\rm ~~~~}\hat{\mathcal{S}}_-(N_0)=\ket{N_0-1}\bra{N_0}\\ 
&\hat{\mathcal{P}}_+(N_0) = \ket{N_{0}+2}\bra{N_0}, {\rm ~~~~}\hat{\mathcal{P}}_-(N_0) = \ket{N_{0}-2}\bra{N_0}\ .
\end{split}
\label{jumps}
\end{equation}
\end{widetext}
Obviously, $\hat{\mathcal{S}}_+(N_0)$ adds one particle to the condensate with a rate 
$\xi_N^+(N_0,T)=2(N_0+1)\lambda^{+}_{\rightsquigarrow}(N-N_0,T)$, 
whereas $\hat{\mathcal{S}}_-(N_0)$ destroys a condensate particle with the rate 
$\xi_N^{-}(N_0,T) = 2N_0\lambda^{-}_{\rightsquigarrow}(N-N_0,T)$, given a number of 
$(N-N_0)$ non-condensate particles, and a temperature $T$ of the heat reservoir.
In a similar way, $\hat{\mathcal{P}}_{\pm}(N_0)$ describe the 
simultaneous creation of two condensate particles 
with a rate $\gamma_N^+(N_0,T)=(N_{0}+1)(N_{0}+2)\lambda^{+}_{\leftrightsquigarrow}(N-N_0,T)$, and 
the depletion of two condensate particles with a 
rate $\gamma_N^-(N_0,T)=N_{0}(N_{0}-1)\lambda^{-}_{\leftrightsquigarrow}(N-N_0,T)$. 
The different transition rates $\xi_N^{\pm}(N_0,T)$ and $\gamma_N^{\pm}(N_0,T)$ are defined by the 
following integrals over non-condensate correlation functions:
\begin{widetext}
\begin{eqnarray}
\lambda^{\pm}_\rightsquigarrow (N-N_0,T) & = & {\rm Re}\left\{ 
\frac{g^{2}}{\hbar^{2}}\iint\limits{\rm d}\vec{\mbf{r}}~{\rm d}\vec{\mbf{r}}^{\prime}~\Psi^{\star}_{0}(\vec{\mbf{r}})\Psi_{0}(\vec{\mbf{r}}^{\prime}) 
\int\limits_{0}^{\infty}{\rm d}\tau~{\rm e}^{-\Gamma^2\tau^2} {\rm e}^{\pm\frac{i\mu_0\tau}{\hbar}}
\mathcal{G}^{(\pm)}_{\rightsquigarrow}(\vec{\mbf{r}},\vec{\mbf{r}}^{\prime},N-N_{0},T,\tau)\right\}
\label{lambda1}\\
\lambda^{\pm}_\leftrightsquigarrow (N-N_0,T) & = & {\rm Re}\left\{\frac{g^{2}}{4\hbar^{2}}\iint{\rm d}\vec{\mbf{r}}~
{\rm d}\vec{\mbf{r}}^{\prime}~\Psi_{0}(\vec{\mbf{r}}) \Psi_{0}(\vec{\mbf{r}}) 
\Psi^{\star}_{0}(\vec{\mbf{r}}^{\prime})\Psi^{\star}_{0}(\vec{\mbf{r}}^{\prime})
\int\limits_{0}^{\infty}{\rm d}\tau~{\rm e}^{-\Gamma^2\tau^2} {\rm e}^{\pm\frac{2i\mu_0\tau}{\hbar}}
\mathcal{G}^{(\pm)}_{\leftrightsquigarrow}(\vec{\mbf{r}},\vec{\mbf{r}}^{\prime},N-N_{0},T,\tau)\right\} 
\label{lambda4}
\end{eqnarray}\\
\end{widetext}
where $\mathcal{G}^{(\pm)}_{\rightsquigarrow}(\vec{\mbf{r}},\vec{\mbf{r}}^{\prime},N-N_{0},T,\tau)$ and 
$\mathcal{G}^{(\pm)}_{\leftrightsquigarrow}(\vec{\mbf{r}},\vec{\mbf{r}}^{\prime},N-N_{0},T,\tau)$ 
are correlation functions of the non-condensate field for 
single particle ($\rightsquigarrow$) and pair ($\leftrightsquigarrow$) events, 
given that $(N-N_0)$ particle are in the non-condensate gas.
In Eqs.~(\ref{lambda1},~\ref{lambda4}), we have extended the time 
integral from $\Delta t$ to $\infty$, assuming a Gaussian decay of 
non-condensate correlations due to thermalization which occurs within a time interval  
on the order of the average time $\Gamma=\tau^{-1}_{{\rm col}}$ for two-body collisions.

The remaining coherent parts $\mathcal{G}^{(\pm)}_{\rightsquigarrow}$ and $\mathcal{G}^{(\pm)}_{\leftrightsquigarrow}$ 
of non-condensate correlations are determined by the thermalized state in Eq.~(\ref{NC_thermal}):
\begin{widetext}
\begin{equation}
\begin{split}
&\mathcal{G}^{(+)}_{\rightsquigarrow}(\vec{\mbf{r}},\vec{\mbf{r}}^{\prime},N-N_{0},T,\tau)  =
\left\langle\hat{\Psi}_{\perp}^{\dagger}(\vec{\mbf{r}},\tau)
\hat{\Psi}_{\perp}^{\dagger}(\vec{\mbf{r}},\tau)\hat{\Psi}_{\perp}(\vec{\mbf{r}},\tau)
\hat{\Psi}_{\perp}^{\dagger}(\vec{\mbf{r}}^{\prime},0)\hat{\Psi}_{\perp}(\vec{\mbf{r}}^{\prime},0)
\hat{\Psi}_{\perp}(\vec{\mbf{r}}^{\prime},0)\right\rangle^{(N-N_{0})}_{\mathcal{F}_\perp}\ , \\  
&\mathcal{G}^{(-)}_{\rightsquigarrow}(\vec{\mbf{r}},\vec{\mbf{r}}^{\prime},N-N_{0},T,\tau) = 
\left\langle\hat{\Psi}_{\perp}^{\dagger}(\vec{\mbf{r}},\tau)
\hat{\Psi}_{\perp}(\vec{\mbf{r}},\tau)\hat{\Psi}_{\perp}(\vec{\mbf{r}},\tau)
\hat{\Psi}_{\perp}^{\dagger}(\vec{\mbf{r}}^{\prime},0)\hat{\Psi}^{\dagger}_{\perp}
(\vec{\mbf{r}}^{\prime},0)\hat{\Psi}_{\perp}(\vec{\mbf{r}}^{\prime},0)
\right\rangle^{(N-N_{0})}_{\mathcal{F}_\perp}\ , \\
\end{split}
\label{twopoint1}
\end{equation} 
for single particle events, and by 
\begin{equation}
\begin{split}
&\mathcal{G}^{(+)}_{\leftrightsquigarrow}(\vec{\mbf{r}},\vec{\mbf{r}}^{\prime},N-N_{0},T,\tau) = 
\left\langle\hat{\Psi}_{\perp}^{\dagger}(\vec{\mbf{r}},\tau)\hat{\Psi}_{\perp}^{\dagger}(\vec{\mbf{r}},\tau)
\hat{\Psi}_{\perp}(\vec{\mbf{r}}^{\prime},0)\hat{\Psi}_{\perp}(\vec{\mbf{r}}^{\prime},0)
\right\rangle_{\mathcal{F}_\perp}^{(N-N_{0})} \ , \\
&\mathcal{G}^{(-)}_{\leftrightsquigarrow}(\vec{\mbf{r}},\vec{\mbf{r}}^{\prime},N-N_{0},T,\tau) = 
 \left\langle\hat{\Psi}_{\perp}(\vec{\mbf{r}},\tau)\hat{\Psi}_{\perp}(\vec{\mbf{r}},\tau)
\hat{\Psi}^{\dagger}_{\perp}(\vec{\mbf{r}}^{\prime},0)\hat{\Psi}^{\dagger}_{\perp}
(\vec{\mbf{r}}^{\prime},0)\right\rangle_{\mathcal{F}_\perp}^{(N-N_{0})}\ , \\
\end{split}
\label{twopoint2}
\end{equation} 
\end{widetext}
for pair events. In Eqs.~(\ref{twopoint1},~\ref{twopoint2}), $\langle\ldots\rangle_{(N-N_0)}$ denotes the average 
${\rm Tr}_{{\mathcal F}_\perp}\left\{\dots \hat{\rho}_\perp(N-N_0,T)\right\}$ 
with respect to a thermal non-condensate state with $(N-N_0)$ particles. 
Note that the imaginary parts of  Eqs.~(\ref{lambda1}-\ref{lambda4}) which lead, in principle, to a shift of the condensate energy levels
(similar to the Lamb shift known from quantum electrodynamics~\cite{Tan_Dup_Gry}), drop out from the master equation
due to the diagonal form of the reduced density matrix, Eq.~(\ref{state_C}).\\

\subsubsection{Quantum master equation of Bose-Einstein condensation}
From the master equation of Lindblad type in Eq.~(\ref{Lindblad_final}), 
we can derive the evolution equation for the condensate particle number distribution, 
$p_{N}(N_{0},t)=\bra{N_{0}}\hat{\rho}^{(N)}_{0}(t)\ket{N_{0}}$.
Considering only single particle processes ($\rightsquigarrow$), since 
they dominate the condensation process in three-dimensional harmonic traps, 
see section~\ref{rates}, leads to 
the quantum master equation for quantum jump processes with $\Delta N_0=\pm1$:
\begin{equation}
\begin{split}
\frac{\partial p_{N}(N_{0},t)}{\partial t} =&- \left[\xi^{+}_{N}\left(N_{0},T\right)+
\xi^{-}_{N}(N_{0},T)\right]p_{N}(N_{0},t) \\
&+ \xi^{+}_{N}(N_{0}-1,T)p_{N}(N_{0}-1,t) \\
& + \xi^{-}_{N}(N_{0}+1,T)p_{N}(N_{0}+1,t) \ ,  
\end{split}
\label{prop_evolution}
\end{equation}
with $\xi^{+}_{N}(N_{0},T)=2(N_{0}+1)\lambda_{\rightsquigarrow}^{+}(N-N_{0},T)$, 
and $\xi^{-}_{N}(N_{0},T)=2N_{0}\lambda_{\rightsquigarrow}^{-}(N-N_{0},T)$, where the transition rates
$\lambda_{\rightsquigarrow}^{\pm}(N-N_{0},T)$ are given by Eq.~(\ref{lambda1}). 

\begin{figure}[t]
\includegraphics[width=7.0cm, height=5.0cm,angle = 0.0]{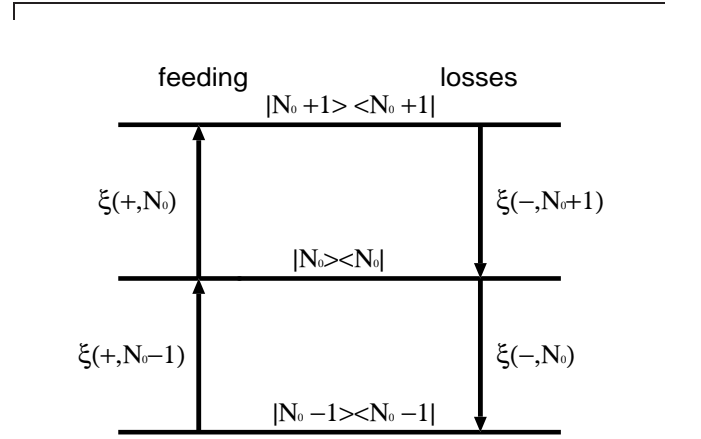}
\caption{Probability flow between different 
condensate number states as expressed 
by the quantum master equation in Eq.~(\ref{prop_evolution}). 
The corresponding transition rates are defined by 
$\xi^{+}_{N}(N_{0},T)=2(N_{0}+1)\lambda_{\rightsquigarrow}^{+}(N_{0},T)$,
and $\xi^{-}_{N}(N_{0},T)=2N_{0}\lambda_{\rightsquigarrow}^{-}(N_{0},T)$, with 
$\lambda_{\rightsquigarrow}^{\pm}(N_{0},T)$ given by Eq.~(\ref{lambda1}). 
In the stationary state which is reached for long times $t\rightarrow\infty$, 
the rates obey the condition of detailed balance: $\xi^{+}_{N}(N_{0},T)
p_N(N_0,T)=\xi^{-}_{N}(N_{0}+1,T)p_N(N_0+1,T)$.} 
\label{probflow}
\end{figure}

Bose-Einstein condensation is now reduced to a simple rate equation, the master Eq.~(\ref{prop_evolution}), 
which describes in particular the buildup of a macroscopic condensate occupation from the fluctuating 
thermal vapor.
As sketched in Fig.~\ref{probflow}, net particle flow \textit{towards} a state $\ket{N_0}\bra{N_0}$
is described 
%in terms of the 
by the
%tw positive probability feeding 
current $\xi_{N}^{+}(N_{0}-1,T)p_N(N_0-1,t) 
+\xi_{N}^{-}(N_{0}+1,T)p_N(N_0+1,t)$,  
%tw whereas 
and
particle flow \textit{from} the state $\ket{N_0}\bra{N_0}$
by the 
%tw negative probability loss
%tw [According to Dominique's comment, "feeding" and "loss" should be related to xi^+ and xi^- rates]
current 
 $\xi_{N}^{+}(N_{0},T)p_N(N_0,t) +
\xi_{N}^{-}(N_{0},T)p_N(N_0,t)$.
As will be shown in section \ref{section_steady}, the steady state of the system is 
therefore reached, if, and only if the net probability flux for every 
state $\ket{N_0}\bra{N_0}$ ($\rightarrow$ detailed balance of probability flow) is zero, i.e. $\xi^{+}_{N}(N_{0},T)
p_N(N_0,T)=\xi^{-}_{N}(N_{0}+1,T)p_N(N_0+1,T)$ for all $N_0=0\ldots N$. 

\subsubsection{Transition rates for Lindblad dynamics}
\label{rates}
We now evaluate the different decay rates, Eqs.~(\ref{lambda1}-\ref{lambda4}). 
For this purpose, we decompose the higher order
correlation functions of the non-condensate fields into second order correlation functions 
according to the Wick theorem~\cite{Wick2,Wick3}, and perform 
the integrals over $\vec{\mbf r}$, $\vec{\mbf r'}$ and $\tau$, see appendix~\ref{appendix_one}.
For the single particle creation and loss events, the result is: 
\begin{equation}
\begin{split}
\lambda^{\rightsquigarrow}_{\pm}(N_\perp,T)=&\frac{16\pi^{3}
\hbar^{2}a^{2}}{m^{2}}\sum_{k,l,m\ne0}\Bigl[f^{\rightsquigarrow}_{\pm}
(k,l,m,N_\perp,T)\times\Bigr.\\
&\times\delta^{{\rm (\Gamma)}}(\omega_{k}+\omega_{l}
-\omega_{m}-\omega_{0})+\\
& \Bigl.+ g^{\rightsquigarrow}_{\pm}
(k,l,m,N_\perp,T)\delta^{{\rm (\Gamma)}}(\omega_{l}-\omega_{0})\Bigr]\\
\end{split}
\label{rate1}
\end{equation}
where $\omega_k \equiv \epsilon_k/\hbar$, $\omega_0\equiv\mu_0/\hbar$, and where
\begin{equation}
\begin{split}
f^{\rightsquigarrow}_{+}(k,l,m,N_{\perp},T) 
&= \langle N_{k}\rangle(N_{\perp},T)\langle N_{l}\rangle(N_{\perp},T)\times\\
&\times\left[\langle N_{m}\rangle(N_{\perp},T)+1\right]\vert\zeta_{kl}^{m0}\vert^{2}\label{fplus}
\end{split}
\end{equation}
and
\begin{equation}
\begin{split}
g^{\rightsquigarrow}_{+}(k,l,m,N_{\perp},T) 
&= 2\langle N_{k}\rangle(N_{\perp},T)\langle N_{l}\rangle(N_{\perp},T)\times\\
&\times\langle N_{m}\rangle(N_{\perp},T) \zeta^{0k}_{kl} \zeta^{ml}_{0m}\label{gplus}
\end{split}
\end{equation}
are the weight functions for condensate particle feedings. Correspondingly,  
\begin{equation}
\begin{split}
f^{\rightsquigarrow}_{-}&(k,l,m,N_{\perp},T) = 
\left[\langle N_{k}\rangle(N_{\perp},T)+1\right]\times\\
&\times\left[\langle N_{l}\rangle(N_{\perp},T)+1\right] 
\langle N_{m}\rangle(N_{\perp},T)\vert\zeta_{kl}^{m0}\vert^{2}\label{fminus}
\end{split}
\end{equation}
and
\begin{equation}
\begin{split}
g^{\rightsquigarrow}_{-}&(k,l,m,N_{\perp},T) = 
2 \langle N_{k}\rangle(N_{\perp},T)\times\\
&\times\left[\langle N_{l}\rangle(N_{\perp},T)+1\right] 
\langle N_{m}\rangle(N_{\perp},T)\zeta^{0k}_{kl} \zeta^{ml}_{0m}\label{gminus}
\end{split}
\end{equation}
are the weight functions for condensate particle losses.
The functions $f_{\rightsquigarrow}^{\pm}(k,l,m,N_{\perp},T)$ and $g_{\rightsquigarrow}^{\pm}(k,l,m,N_{\perp},T)$ depend on temperature, 
the number of non-condensate particles, $N_{\perp}=(N-N_{0})$, 
and on the quantum mechanical probability amplitudes 
\begin{equation}
\zeta_{0m}^{lk}=(\zeta_{kl}^{m0})^{\star} = \int\limits\,{\rm d}\vec{\mbf{r}}~
\Psi^{\star}_{0}(\vec{\mbf{r}})\Psi^{\star}_{m}(\vec{\mbf{r}})
\Psi_{k}(\vec{\mbf{r}})
\Psi_{l}(\vec{\mbf{r}})\ 
\label{overlap}
\end{equation}
for the different microscopic single particle feeding and loss processes 
with energy balances $\Delta\omega_\rightsquigarrow = \Delta\epsilon_{\rightsquigarrow}/\hbar=
(\omega_{k}+\omega_{l}-\omega_{m}-\omega_{0})$ or $(\omega_{l}-\omega_{0})$.
The $\delta$-distribution in Eq.~(\ref{rate1}), $\delta^{{\rm (\Gamma)}}(\Delta\omega_\rightsquigarrow) = 
\sqrt{\pi}/\Gamma^2{\rm exp}[-(\Delta\omega_\rightsquigarrow)^2/4\Gamma^2]$,
reflects conservation of energy during the different single particle feeding and loss processes on a certain width 
$\sqrt{2}\Gamma$ arising from the decay of the 
non-condensate field correlation functions in Eqs.~(\ref{lambda1},~\ref{lambda4}).
Therefore, only single particle processes with 
energy balances $\Delta\epsilon_{\rightsquigarrow}/\hbar<\Gamma$
 will contribute to the rates 
in Eq.~(\ref{rate1}).

The average occupation number $\langle N_{k}\rangle(N_\perp,T)$ 
of a non-condensate single particle state $\ket{\Psi_{k}}$, 
given a thermal state projected onto the subspace of $N_{\perp}=(N-N_{0})$ 
non-condensate atoms, reads (see appendix~\ref{appendix_occs})
\begin{equation}
\langle N_{k}\rangle(N_{\perp},T)=\frac{1}{{\rm exp}[\beta(\epsilon_{k}-\mu_{\perp}(N_\perp,T)]-1}\ ,
\label{occupation}
\end{equation}
where $\mu_{\perp}(N_{\perp},T)$ is defined by the normalization condition:
$\sum_{k\ne0}\langle N_{k}\rangle(N_{\perp},T)=N_\perp$. According to this definition,
$\mu_{\perp}(N_{\perp},T)$ equals the chemical potential of a thermal state of 
$N_\perp=(N-N_0)$ non-condensate particles~\cite{Statistics}.
From $\langle N_k\rangle+1=\langle N_k\rangle \exp[\beta(\epsilon_k-\mu_\perp)]$, 
and using the energy conservation as expressed by the $\delta$-function in Eq.~(\ref{rate1}), 
one can derive the following relation between the single particle loss and feeding rates:
\begin{equation}
 \lambda^{+}_{\rightsquigarrow}(N-N_{0},T) = 
 {\rm e}^{\beta\Delta\mu(N-N_{0},T)}
 \lambda^{-}_{\rightsquigarrow}(N-N_{0},T)\ ,
\label{detailed_balance}
\end{equation}
where $\Delta\mu(N-N_{0})=\mu_\perp(N-N_0,T)-\mu_{0}$. To obtain Eq.~(\ref{detailed_balance}), the 
finite width $\Gamma$ of the $\delta$-function is neglected, which is justified under the condition $\hbar\Gamma\beta\ll1$. 
The relation (\ref{detailed_balance}) will be useful in Sec.~\ref{section_steady} to determine the equilibrium state
of the Bose gas. 

Note that Eq.~(\ref{occupation}) takes into account the depletion of the non-condensate during condensate formation,
ensuring that $\langle N_k\rangle\to 0$ as $N_0\to N$. According to Eqs.~(\ref{fplus},\ref{fminus}), also the condensate feeding and loss rates tend to zero in this limit. In contrast, the rates obtained within quantum kinetic theory \cite{QKT} increase with increasing condensate particle number $N_0$. 

Finally, the rates for pair events turn into:
\begin{equation}
\begin{split}
\lambda^{\leftrightsquigarrow}_{\pm}(N_{\perp},T)
=&\frac{4\pi^{3}\hbar^{2}a^{2}}{m^{2}}
\sum_{k,l\ne0}f^{\leftrightsquigarrow}_{\pm}
(k,l,N_{\perp},T)\\
&\times\delta^{{\rm (\Gamma)}}(\omega_{k}+
\omega_{l}-2\omega_{0})\ ,\\
\end{split}
\label{rate2}
\end{equation}
with the weight function 
\begin{equation}
f^{\leftrightsquigarrow}_{+}(k,l,N_{\perp},T) 
= \langle N_{k}\rangle(N_{\perp},T)\langle N_{l}\rangle(N_{\perp},T)\vert\zeta_{kl}^{00}\vert^{2}
\end{equation}
for pair feedings, and correspondingly 
\begin{equation}
\begin{split}
f^{\leftrightsquigarrow}_{-}&(k,l,N_{\perp},T) 
= (\langle N_{k}\rangle(N_{\perp},T)+1)\times\\
&\times(\langle N_{l}\rangle(N_{\perp},T)+1)
\vert\zeta_{kl}^{00}\vert^{2}\\
\end{split}
\end{equation}
for pair losses, with 
\begin{equation}
\zeta_{00}^{kl} 
= \int\,{\rm d}\vec{\mbf{r}}~
(\Psi^{2}_{0}(\vec{\mbf{r}}))^{\star}
\Psi_{k}(\vec{\mbf{r}})
\Psi_{l}(\vec{\mbf{r}})\ .
\end{equation} 
Looking at the energy balance of a pair 
event, $\Delta\epsilon_{\leftrightsquigarrow}/\hbar=\omega_{k}
+\omega_{l}-2\omega_{0}$, we see that pair events 
occur as energy non-conserving processes, i.e.,  $\Delta\epsilon_{\leftrightsquigarrow}/\hbar \gg \Gamma\sim10-20$ Hz, 
since the single particle energy $\mu_0$ of condensate particles is 
smaller than the energies $\epsilon_{k,l}\gg\hbar\Gamma$
 of non-condensate particles 
in a three-dimensonal harmonic trap.
Therefore, pair events can be neglected in comparison with the single-particle events in the master equation (\ref{prop_evolution}).
For the same reason, we can neglect the terms associated to the $g^{\rightsquigarrow}_{\pm}$ functions in Eq.~(\ref{rate1}) as compared to those associated to the $f^{\rightsquigarrow}_{\pm}$ functions.

\section{Results}
\label{sectionV}
\label{section_times}
\begin{figure}[t]
\includegraphics[width=8.5cm, height=5.2cm,angle = 0.0]{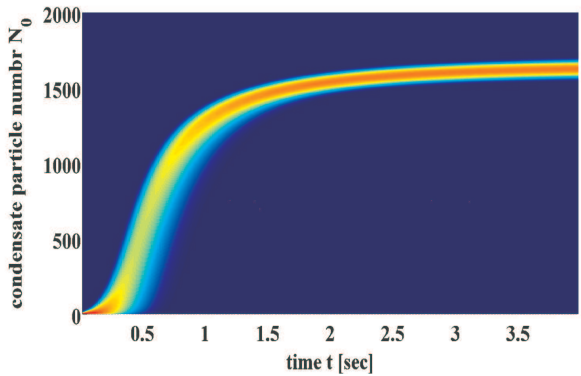}
\caption{(color online) Time evolution of $p_N(N_0,t)$ (probability is 
indicated by the color gradient in the figure) according to Eq.~(\ref{prop_evolution}) 
during condensate formation for parameters similar to Ref. \cite{BEC_exp}, i.e. 
in a gas of $N=2000$ $^{87}$Rb atoms confined a harmonic trap with frequencies 
$\omega_x=\omega_y=2\pi\times42.0$ Hz and $\omega_z=2\pi\times120.0$ Hz. 
The temperature has been set to $T=20.31$ nK, and 
the critical temperature of the gas is $T_{{\rm c}}=33.86$ nK. According to the atomic density 
$\varrho=2.6\times10^{12} {\rm ~cm}^{-3}$ and the s-wave scattering length $a=5.7{\rm ~nm}$, we set 
$\Gamma=(a^2\varrho)^{-1}v=34$ Hz, where $v=\sqrt{3k_BT/m}$. }
\label{fig_formjila}
\end{figure}
In this section, we present numerical studies of the condensate 
particle number distribution obtained from Eq.~(\ref{prop_evolution}) 
during Bose-Einstein condensation, and derive
the unique equilibrium steady state of the Bose-Einstein condensate. Under the assumption $\hbar\Gamma\beta\ll 1$, 
the equilibrium steady state is proven to be a Gibbs-Boltzmann (thermal) state of non-interacting particles 
in the 
dilute regime 
$a N/L\ll 1$, with $L=\hbar/m\omega$ the extension of the harmonic oscillator 
ground state.

\subsection{Perturbative calculation of transition rates}
\label{section_pert}
For this purpose,
we now consider the 
case of very dilute, weakly interacting gases.
Since the transition rates derived in the previous section originate from processes 
of second order in the interaction $\hat{\mathcal V}_{\perp 0}$,
all the rates are proportional to $a^2$, as evident from the prefactors in Eqs.~(\ref{rate1},~\ref{rate2}). 
The remaining dependence of the rates on the interaction strength originates from the single particle 
wave functions $|\Psi_k\rangle$, the eigenvalue of the Gross-Pitaevskii equation $\mu_0$, 
and the non-condensate single particle energies $\epsilon_k$, which are themselves functions on the 
parameter $a\varrho$, see Eq.~(\ref{timeindependent_GP}).

Interested in the 
case of dilute and weakly interacting gases,
we thus expand these quantities (i.e. $|\Psi_0\rangle$, $\mu_0$ and $\epsilon_k$) in terms of the 
scattering length $a$,
taking into account only the first non-vanishing contribution, given 
by their non-interacting limits.
Hence, the basis states $\ket{\Psi_k}$ turn into the single particle eigenstates $\ket{\chi_k}$ 
of the trapping potential, with corresponding ground state energy 
$\epsilon_0$, whereas the $\epsilon_k 's$ are the energies of the excited states. 
Thereby, we evaluate the transition rates to lowest non-vanishing order -- proportional to 
$a^2$ -- in the s-wave scattering length.

Quantitatively, this procedure is correct as long as the ground state of the Gross-Pitaevski equation (\ref{timeindependent_GP})
is well approximated by the single-particle ground state. This, in turn, is the case if the interaction energy $gN|\Psi_0|^2$ is much smaller than the 
harmonic oscillator energy $\hbar\omega$, or, in other words, if 
$aN/L\ll1$, where 
$L=\hbar/m\omega$ denotes the extension of the harmonic oscillator 
ground state.

When $aN/L$ increases, the ground state of the Gross-Pitaevskii equation is
progressively distorted, as well as the various excited states, making the
explicit calculation of the rates, Eqs.(\ref{rate1}-\ref{overlap}), more difficult. However,
no drastic change is likely to take place, making the following predictions
qualitatively and maybe semi-quantitatively correct for a realistic situation
like Bose-Einstein condensation of a Rb or Na gas.

\subsection{Dynamics of Bose-Einstein condensation}
\label{sec_dyn}
Equation~(\ref{prop_evolution}) is solved numerically 
to propagate the condensate particle number distribution $p_N(N_0,t)$ in time.
Figure~\ref{fig_formjila} displays a typical example for the time evolution 
of $p_N(N_0,t)$ for a gas of $N=2000$ $^{87}$Rb atoms which undergoes 
the Bose-Einstein condensation phase transition 
in a three-dimensional harmonic trap with 
frequencies $\omega_x=\omega_y=2\pi\times42.0$ Hz, $\omega_z=2\pi\times120.0$ Hz. 
The final temperature of the gas is $T=20.31$ nK, given an ideal gas critical 
temperature $T_{{\rm c}}=33.86$ nK~\cite{Str_Pit_BEC}.
Note that, $\Gamma$ is a free parameter in our theory, provided it is 
smaller than $k_BT$ and larger than the external trap frequency. However, 
we have numerically checked that the transition rates do not significantly change 
with varying $\Gamma$ in this
%tw the said
 parameter regime. 

To calculate the transition rates leading to the condensate growth scenario in Fig.~\ref{fig_formjila}, we used the semi-classical limit 
\cite{Str_Pit_BEC, Schelle}, where the discrete sums in the feeding and loss rates in Eq.~(\ref{rate1}) are replaced by an integral over the 
density of states $g(\epsilon)=\epsilon^{2}/2(\hbar^3\omega_x\omega_y\omega_z)$.
This shifts the final condensate 
fraction by appr.
$10\%$ as compared to the exact numerical 
evaluation of the discrete sums (employed in Figs.~\ref{fig_ptndist} and \ref{comp3} below), but does not change the qualitative behavior observed in Fig.~\ref{fig_formjila}.

\subsection{Average condensate growth}
\label{section_times2}
From the time evolution of the distribution $p_N(N_0,t)$ 
defined by Eq.~(\ref{prop_evolution}) the 
growth of the average condensate population can be extracted using
\begin{equation}
\langle N_0\rangle(t) = \sum_{N_0=0}^N N_0 p_N(N_0,t)\ ,
\end{equation} 
and deriving a corresponding condensate growth equation~\cite{QKT}.
For this purpose, we assume a sufficiently narrow peaked distribution 
$p_N(N_0,t)$ around the mean value $\langle N_0\rangle$ as indicated by Fig.~\ref{fig_formjila}, 
such that the rates are approximately constant in this narrow region, meaning that
\begin{equation}
\lambda^{\pm}_{\rightsquigarrow}(N-N_{0},T)\approx \lambda^{\pm}_{\rightsquigarrow}(N-\langle N_{0}\rangle,T)\ ,
\label{approx1}
\end{equation}
for $N_0$ close enough to $\langle N_0\rangle$. Taking the average $N_0\partial p(N_0,t)/\partial t$ with 
$\partial p(N_0,t)/\partial t$ given by Eq.~(\ref{prop_evolution}) finally leads to the 
following growth equation for the 
average condensate occupation:
\begin{equation}
\frac{\partial \langle N_{0}\rangle}{\partial t}=\xi^{+}_{N}(\langle N_{0}\rangle,T)
-\xi^{-}_{N}(\langle N_{0}\rangle+1,T)\ ,
\label{av_N_easy}
\end{equation}
with $\xi^{+}_{N}(\langle N_{0}\rangle,T)=2(\langle N_{0}\rangle+1)\lambda_{\rightsquigarrow}^{+}(N-\langle N_{0}\rangle,T)$, 
and $\xi^{-}_{N}(\langle N_{0}\rangle+1,T)=2(\langle N_{0}\rangle+1)\lambda_{\rightsquigarrow}^{-}(N-\langle N_{0}\rangle,T)$.
The equilibrium value of $\langle N_0\rangle$ is hence defined by the detailed particle balance condition
$\lambda^{+}_{\rightsquigarrow}(N-\langle N_{0}\rangle,T)=\lambda^{-}_\rightsquigarrow(N-\langle N_{0}\rangle,T)$.
According to the above relation between the rates $\lambda^{+}_{\rightsquigarrow}$ and $\lambda^{-}_{\rightsquigarrow}$, 
see Eq.~(\ref{detailed_balance}), this implies equality of the chemical potentials on average:
\begin{equation}
\mu_0=\mu_\perp(N-\langle N_0\rangle,T)\ .
\end{equation}
In the next section, we will show that not only the average condensate occupation, 
but also the whole steady state distribution $p_N(N_0,T)$
agrees with the 
thermodynamical prediction.

\subsection{Steady state distribution}
\label{section_steady}
\begin{figure}[t]
\begin{center}
\includegraphics[width=8.2cm,height=5.4cm,angle = 0.0]{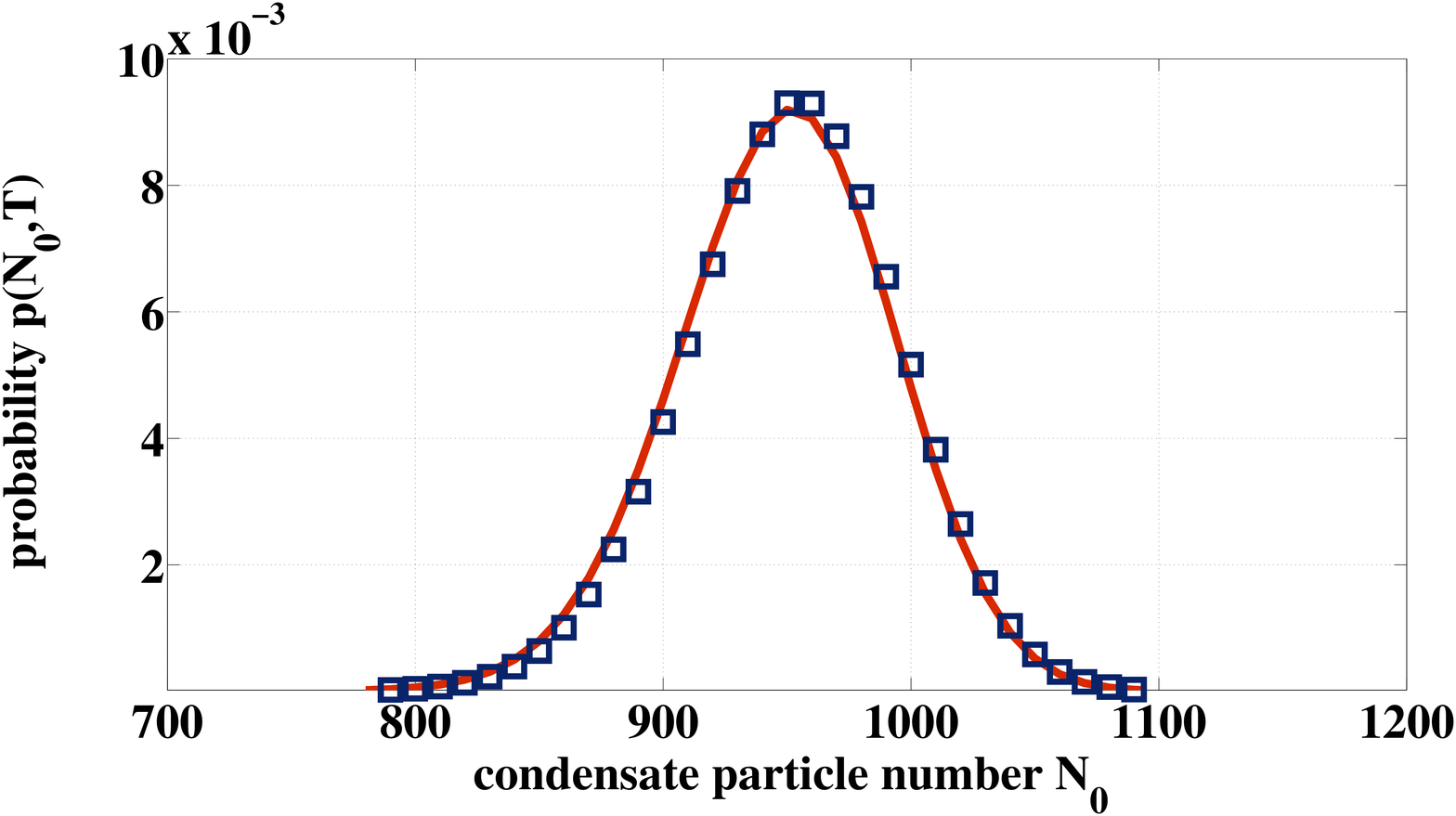}
\caption{(color online) Comparison of condensate particle number 
distribution arising from the master equation (red solid line) vs. the 
Boltzmann thermal state ansatz of Eq.~(\ref{Boltzmann}) (blue squares), for 
the same parameters as in Fig.~\ref{fig_formjila}. The gas temperature is $T=25.0$ nK, 
i.e. $\beta\hbar\Gamma \sim 0.01$, with $\Gamma=34$ Hz.}
\label{fig_ptndist}
\end{center}
\end{figure}

\begin{figure}[t]
\begin{center}
\includegraphics[width=7.3cm,height=4.5cm,angle = 0.0]{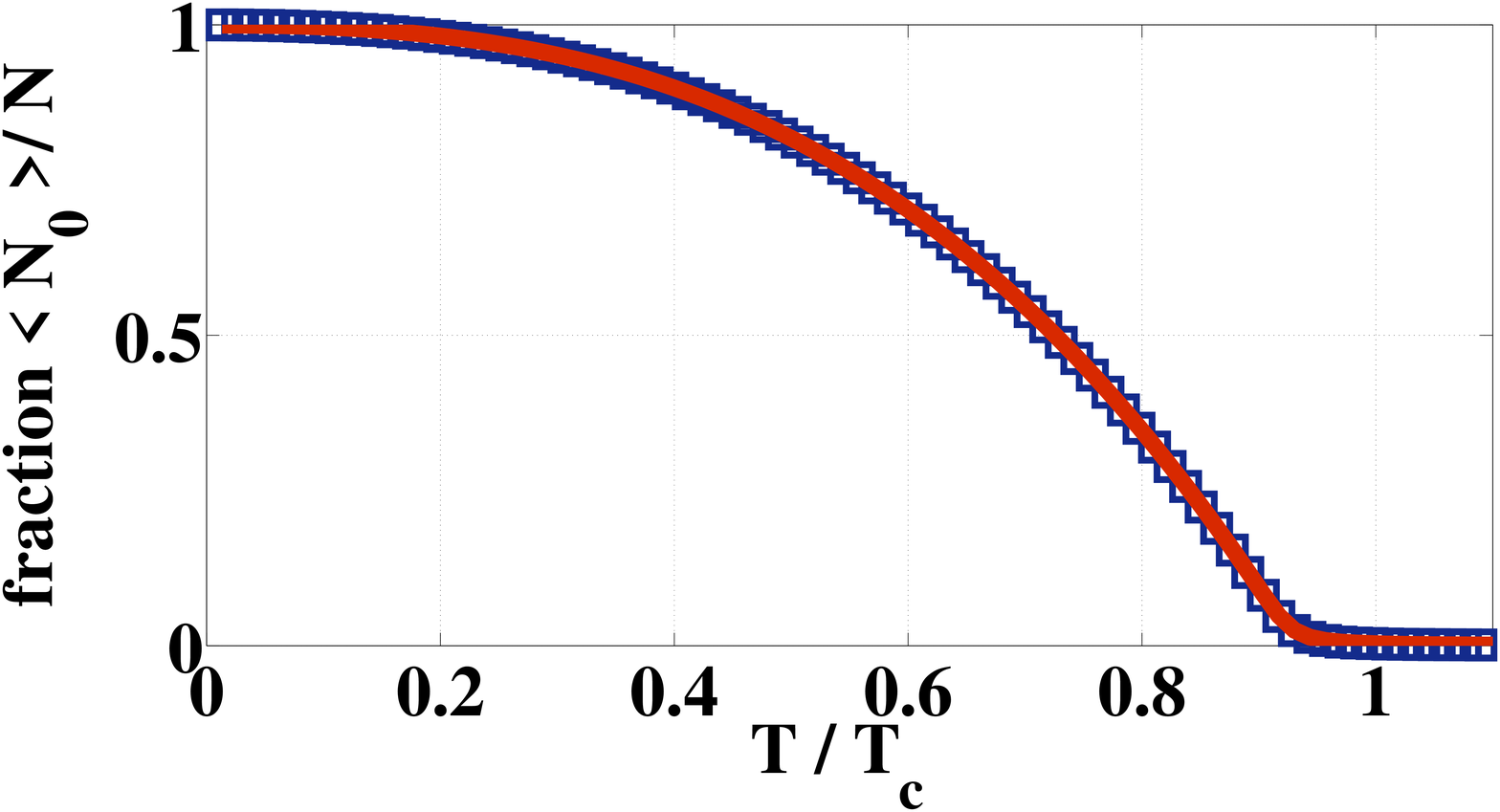}
\includegraphics[width=7.3cm,height=4.5cm,angle = 0.0]{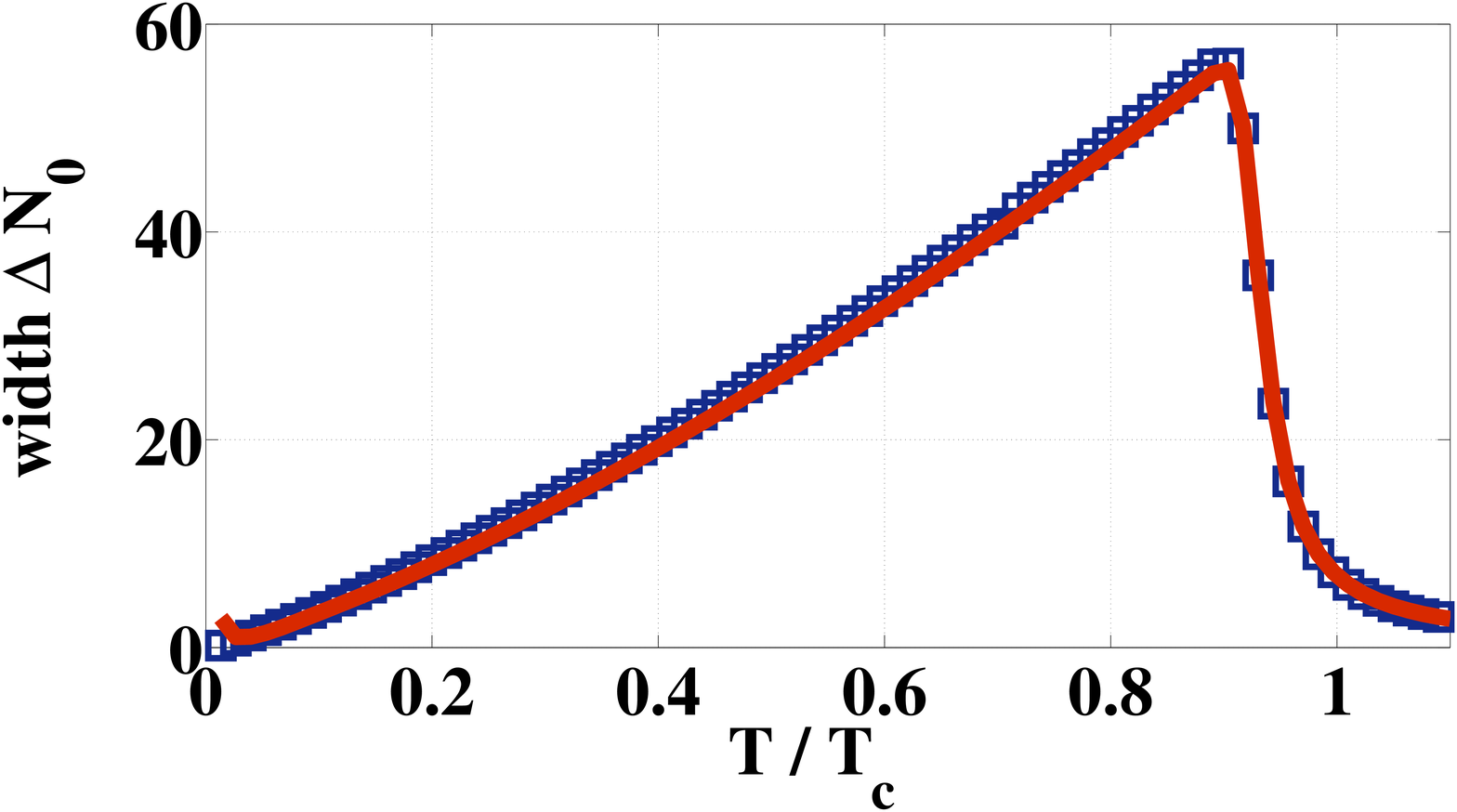}\\
\caption{(color online) Average condensate fraction $\langle N_0\rangle(t) / N$ (upper panel), 
and standard deviation $\Delta N_0$ (lower panel) of the condensate particle number distribution $p_N(N_0,T)$
as a function of temperature, and otherwise the same parameters as in Fig.~\ref{fig_ptndist}.
 The master equation's steady state distribution (red solid line) agrees very well with
the canonical ensemble prediction (blue squares) of Eq.~(\ref{Boltzmann}).}
\label{comp3}
\end{center}
\end{figure}
For this purpose, we solve Eq.~(\ref{Bose_state_final}) 
for the steady state distribution, defined by $\partial p_N(N_0,t)/\partial t=0$, which leads to: 
\begin{equation}
p_{N}(N_{0},T)=\mathcal{N}\prod_{z=1}^{N_{0}-1}
\frac{\xi^{+}_{N}(z-1,T)}{\xi^{-}_{N}(z,T)} \ .
\label{assumption2}
\end{equation}
Let us now compare this steady state to a thermal state of $N$ 
non-interacting particles at temperature $T$:
\begin{equation}
\hat{\sigma}_{N,\rm th} = \hat{\mathcal{Q}}_N\frac{{\rm e}^{-\beta\hat{\mathcal{H}}}}{\mathcal{Z}(N,T)}\hat{\mathcal{Q}}_N \ ,
\label{Boltzmann}
\end{equation} 
with the partition function $\mathcal{Z}(N,T)$ of $N$ indistinguishable particles, and the projector 
$\hat{\mathcal{Q}}_N$ onto the Fock space of $N$ particles. In the absence of interactions,
$\mathcal{\hat{H}}$ in Eq.~(\ref{Boltzmann}) is the Hamiltonian 
of the gas in Eq.~(\ref{Hamiltonian_full}), 
with $g\equiv0$. 
To proof the equality of the state $\hat{\sigma}_{N,\rm th}$ and the steady state 
of the Bose gas in Eq.~(\ref{Bose_state_final}), with $p_N(N_0,T)$ given by Eq.~(\ref{assumption2}), 
it needs to be shown that the recursion relation for the condensate particle number 
distribution, 
\begin{equation}
\frac{p_{N,{\rm th}}(N_0,T)}{p_{N,{\rm th}}(N_0+1,T)} = {\rm e}^{\beta\epsilon_{0}}\frac{\mathcal{Z}_\perp(N-N_0,T)}{\mathcal{Z}_\perp(N-N_0-1,T)}\ ,
\label{recurrence_C1}
\end{equation}
which results from tracing Eq.~(\ref{Boltzmann}) over the non-condensate, 
applies as well for the steady state, Eq.~(\ref{assumption2}), of the master equation. 
In Eq.~(\ref{recurrence_C1}), $\mathcal{Z}_\perp(N-N_0,T)$ is 
the partition function of $(N-N_0)$ non-condensate particles, 
see Eq.~(\ref{partition}), and $\epsilon_{0}$ the single particle 
ground state energy of a non-interacting gas.
This can be seen if we approximate
$\lambda^{-}_{\rightsquigarrow}(N-N_{0},T)\approx 
\lambda^{-}_{\rightsquigarrow}(N-N_{0}+1,T)$, neglecting terms 
of the order of $N^{-1}$. In this case, we obtain from 
Eq.~(\ref{assumption2}):
\begin{equation}
\frac{p_{N}(N_0,T)}{p_{N}(N_0+1,T)}\simeq 
\frac{\lambda^{-}_{\rightsquigarrow}(N-N_0,T)}{\lambda^{+}_{\rightsquigarrow}(N-N_0,T)} =
e^{\beta(\epsilon_0-\mu_\perp(N_\perp,T))},
\label{steady}
\end{equation}
where we used Eq.~(\ref{detailed_balance}), 
and $\epsilon_0=\mu_0$ (in the 
regime of small interactions, 
$aN/L\ll 1$,
see Sec.~\ref{section_pert}).

Now, the non-condensate chemical potential, as defined 
by the normalization condition in Eq.~(\ref{occupation}),
can be shown~\cite{Statistics} to be related to the 
non-condensate partition function, via
$-\beta \mu_\perp(N-N_0,T)={\rm ln}\mathcal{Z}_\perp(N-N_0,T) - {\rm ln}\mathcal{Z}_\perp(N-N_0-1,T)$.
Therewith, we arrive at the recurrence 
relation in Eq.~(\ref{recurrence_C1}), which was to be shown. 
Hence, the steady state of the entire Bose 
gas in Eq.~(\ref{Bose_state_final}) is given by the thermal state 
in Eq.~(\ref{Boltzmann}), in the 
case of 
weak interactions, 
and under the condition $\beta\hbar\Gamma\ll 1$ for which Eq.~(\ref{detailed_balance}) 
is proven to be valid.

The $1/N$ approximation required for the above proof is confirmed by comparing
the exact numerical calculation of the steady state 
condensate particle number distribution
to the prediction of the Boltzmann ansatz in Eqs.~(\ref{Boltzmann}). 
Fig.~\ref{fig_ptndist} shows the stationary particle number distribution for the same parameters as in Fig.~\ref{fig_formjila}.
In order to show that the agreement holds up to the critical temperature (and beyond),
Fig.~\ref{comp3} displays
the comparison of 
average condensate occupations, and the standard deviation 
of the stationary condensate particle number distributions (such as the one depicted in Fig.~\ref{fig_ptndist}), 
as a function of the entire range of relative temperatures, $T/T_{{\rm c}}$,
for $N=2000$ atoms for the same trap parameters as in Fig.~\ref{fig_ptndist}.
Again, we observe close agreement between 
master equation and the Boltzmann ansatz: 
The shift of the critical temperature is about 
10\% with respect to the critical temperature $T_{{\rm c}}$ 
of Bose-Einstein statistics in the semiclassical limit~\cite{Str_Pit_BEC} in both cases. 

\section{Conclusion}
\label{sectionVI}
We have presented the conceptual part and first numerical results 
of a number-conserving quantum master equation theory 
to describe the 
transition of a 
dilute
gas of 
$N$ bosonic atoms into a 
Bose-Einstein condensate. The central result of our theory is the 
quantum master equation in Eq.~(\ref{prop_evolution}) 
which describes the time evolution of the 
reduced condensate state 
in contact with the non-condensate environment for a fixed total 
atom number.
In the dilute gas regime,
we numerically monitored the full 
condensate particle number
distribution $p_N(N_0,t)$ during condensate formation. 

The theory predicts condensate formation times of the order of seconds, 
matching experimentally and theoretically observed times scales~\cite{BEC_exp, BEC_exp1}.
The derived steady state for a dilute, weakly interacting 
Bose-Einstein condensate undergoing Markovian dynamics 
is unique, and proven to obey the same statistics as a Gibbs-Boltzmann 
thermal state of non-interacting particles, in the 
case of weak interactions
$aN/L\ll 1$,
and for the case $\hbar\beta\Gamma\ll 1$ of not too rapidly 
decaying non-condensate correlation functions. 

Future improvements of our model will consist in a microscopic derivation of the
rate $\Gamma$ 
describing the decay
of non-condensate 
field correlation functions (e.g. by diagrammatic expansion techniques for 
higher order correlation functions),
which, in the present version, has been introduced in a rather phenomenological way.
Furthermore, the condition $aN/L\ll 1$ of weak interactions 
may be relaxed. 
Since, in this case, the condensate wave function will depend on the number of condensate particles, this will in particular require to introduce
time-dependent condensate and non-condensate wave functions. 
Finally, it remains to be studied whether 
deviations from the Gibbs-Boltzmann occur if the condition
$\hbar\beta\Gamma\ll 1$ is not fulfilled.
 
A.S. acknowledges financial support 
from QUFAR Marie Curie Action 
MEST-CT-2004-503847, and partial funding through DFG (Forschergruppe 760). 
We thank Boris Fine, Beno\^{i}t Gr\'{e}maud, Markus Oberthaler,
Peter Schlagheck und Alice Sinatra for helpful discussions.  
A.S. is grateful to Cord M\"{u}ller, for hospitality at the University of Bayreuth, 
and for stimulating questions during the development of the theory. 

\appendix
\section{Two point correlation functions} 
\label{appendix_one}
Here, we decompose 
%tw normally and anti-normally ordered two point correlation functions 
the correlation functions of 
of the non-condensate field for single particle ($\rightsquigarrow$) and  
pair ($\leftrightsquigarrow$) processes into products of two-point correlation functions with Wick's theorem, which applies to thermal expectation values \cite{Wick2,Wick3}.

We begin with the 
%tw normally ordered two point 
correlation function for single particle processes
$\mathcal{G}^{(+)}_{\rightsquigarrow}
(\vec{\mbf{r}},\vec{\mbf{r}}^{\prime},N-N_{0},T,\tau)$ in Eq.~(\ref{twopoint1}):
%\vspace{-.8cm}
\begin{widetext}
%\vspace{-.8cm}
\begin{equation}
\begin{split}
&\mathcal{G}^{(+)}_{\rightsquigarrow}
(\vec{\mbf{r}},\vec{\mbf{r}}^{\prime},N-N_{0},T,\tau)=\left\langle\hat{\Psi}_{\perp}^{\dagger}(\vec{\mbf{r}},\tau)
\hat{\Psi}_{\perp}^{\dagger}(\vec{\mbf{r}},\tau)\hat{\Psi}_{\perp}(\vec{\mbf{r}},\tau)
\hat{\Psi}_{\perp}^{\dagger}(\vec{\mbf{r}}^{\prime},0)\hat{\Psi}_{\perp}(\vec{\mbf{r}}^{\prime},0)
\hat{\Psi}_{\perp}(\vec{\mbf{r}}^{\prime},0)\right\rangle^{(N-N_{0})}_{\mathcal{F}_\perp} = \\
&2\left\langle\hat{\Psi}_{\perp}^{\dagger}(\vec{\mbf{r}},\tau)
\hat{\Psi}_{\perp}(\vec{\mbf{r}}^{\prime},0)\right\rangle^{(N-N_{0})}_{\mathcal{F}_\perp}
\left\langle\hat{\Psi}_{\perp}^{\dagger}(\vec{\mbf{r}},\tau)
\hat{\Psi}_{\perp}(\vec{\mbf{r}}^{\prime},0)\right\rangle^{(N-N_{0})}_{\mathcal{F}_\perp}
\left\langle\hat{\Psi}_{\perp}(\vec{\mbf{r}},\tau)
\hat{\Psi}_{\perp}^{\dagger}(\vec{\mbf{r}}^{\prime},0)\right\rangle^{(N-N_{0})}_{\mathcal{F}_\perp} + \\
&4\left\langle\hat{\Psi}_{\perp}^{\dagger}(\vec{\mbf{r}},\tau)
\hat{\Psi}_{\perp}(\vec{\mbf{r}},\tau)\right\rangle^{(N-N_{0})}_{\mathcal{F}_\perp}
\left\langle\hat{\Psi}_{\perp}^{\dagger}(\vec{\mbf{r}},\tau)
\hat{\Psi}_{\perp}(\vec{\mbf{r}}^{\prime},0)\right\rangle^{(N-N_{0})}_{\mathcal{F}_\perp}
\left\langle\hat{\Psi}_{\perp}^{\dagger}(\vec{\mbf{r}}^{\prime},0)\hat{\Psi}_{\perp}(\vec{\mbf{r}}^\prime,0)\right\rangle^{(N-N_{0})}_{\mathcal{F}_\perp}\ .
\end{split}
\label{deco_corr2}
\end{equation}
%tw The anti-normally ordered correlation function for single particle processes
%tw $\mathcal{G}^{(-)}_{\rightsquigarrow}(\vec{\mbf{r}},\vec{\mbf{r}}^{\prime},N-N_{0},T,\tau)$ in Eq.~(\ref{twopoint2}) can be 
%tw decomposed
and similarly:
\begin{equation}
\begin{split}
&\mathcal{G}^{(-)}_{\rightsquigarrow}
(\vec{\mbf{r}},\vec{\mbf{r}}^{\prime},N-N_{0},T,\tau) = \left\langle\hat{\Psi}_{\perp}^{\dagger}(\vec{\mbf{r}},\tau)
\hat{\Psi}_{\perp}(\vec{\mbf{r}},\tau)\hat{\Psi}_{\perp}(\vec{\mbf{r}},\tau)
\hat{\Psi}_{\perp}^{\dagger}(\vec{\mbf{r}}^{\prime},0)\hat{\Psi}^{\dagger}_{\perp}(\vec{\mbf{r}}^{\prime},0)
\hat{\Psi}_{\perp}(\vec{\mbf{r}}^{\prime},0)\right\rangle^{(N-N_{0})}_{\mathcal{F}_\perp} = \\
&2\left\langle\hat{\Psi}_{\perp}^{\dagger}(\vec{\mbf{r}},\tau)
\hat{\Psi}_{\perp}(\vec{\mbf{r}}^{\prime},0)\right\rangle^{(N-N_{0})}_{\mathcal{F}_\perp}
\left\langle\hat{\Psi}_{\perp}(\vec{\mbf{r}},\tau)
\hat{\Psi}^{\dagger}_{\perp}(\vec{\mbf{r}}^{\prime},0)\right\rangle^{(N-N_{0})}_{\mathcal{F}_\perp}
\left\langle\hat{\Psi}_{\perp}(\vec{\mbf{r}},\tau)
\hat{\Psi}_{\perp}^{\dagger}(\vec{\mbf{r}}^{\prime},0)\right\rangle^{(N-N_{0})}_{\mathcal{F}_\perp}+\\
&4\left\langle\hat{\Psi}_{\perp}^{\dagger}(\vec{\mbf{r}},\tau)
\hat{\Psi}_{\perp}(\vec{\mbf{r}},\tau)\right\rangle^{(N-N_{0})}_{\mathcal{F}_\perp}
\left\langle\hat{\Psi}_{\perp}(\vec{\mbf{r}},\tau)
\hat{\Psi}^{\dagger}_{\perp}(\vec{\mbf{r}}^{\prime},0)\right\rangle^{(N-N_{0})}_{\mathcal{F}_\perp}
\left\langle\hat{\Psi}_{\perp}^{\dagger}(\vec{\mbf{r}}^{\prime},0)\hat{\Psi}_{\perp}(\vec{\mbf{r}}^\prime,0)\right\rangle^{(N-N_{0})}_{\mathcal{F}_\perp}\ .
\end{split}
\label{deco_corr1}
\end{equation}
The non-condensate field $\hat{\Psi}_{\perp}(\vec{\mbf{r}},\tau)$ in the interaction picture with respect to 
$\hat{\mathcal{H}}_{\perp}$ in Eq.~(\ref{Hamiltonian_gas}), written in the single particle basis set 
$\lbrace\ket{\Psi_{k}}, k\in\mathbb{N}\rbrace$, turns into 
\begin{equation}
\hat{\Psi}_{\perp}(\vec{\mbf{r}},\tau) = \hat{\mathcal{U}}_{\perp}(\tau) 
\hat{\Psi}_{\perp}(\vec{\mbf{r}})\hat{\mathcal{U}}^{\dagger}_{\perp}(\tau) = 
\sum_{k\ne0}\Psi_{k}(\vec{\mbf{r}})
\hat{a}_{k}{\rm exp}\left[-\frac{i\epsilon_{k}\tau}{\hbar}\right]\ .
\label{sum_IA}
\end{equation}
Any two point correlation function of products of two non-condensate fields in 
Eqs.~(\ref{deco_corr2}) and (\ref{deco_corr1}) can thus be written in terms of the average 
occupation of different non-condensate single particle states $\ket{\Psi_{k}}\in\mathcal{F}_{\perp}$, e.g.:
\begin{equation}
 \left\langle\hat{\Psi}_{\perp}^{\dagger}(\vec{\mbf{r}},\tau)
\hat{\Psi}_{\perp}(\vec{\mbf{r}}^{\prime},0)\right\rangle^{(N-N_{0})}_{\mathcal{F}_\perp} =
\sum_{k\ne0}\Psi_{k}^{\star}(\vec{\mbf{r}})\Psi_{k}(\vec{\mbf{r}}^{\prime})
\langle N_{k}\rangle(N-N_{0},T)~{\rm exp}\left[-\frac{i\epsilon_{k}\tau}{\hbar}\right]\ ,
\label{twopoint_sum}
\end{equation}
where we used that $\left\langle\hat{a}^{\dagger}_{k}\hat{a}_{l}\right\rangle^{(N-N_{0})}_{\mathcal{F}_\perp} = 
\left\langle\hat{a}^{\dagger}_{k}\hat{a}_{k}
\right\rangle^{(N-N_{0})}_{\mathcal{F}_\perp}\delta_{kl}\ 
\equiv \langle N_{k}\rangle(N-N_{0},T)\delta_{kl}$\ .
The function 
\begin{equation}
\langle N_{k}\rangle(N-N_{0},T) = {\rm Tr}_{\mathcal{F}_\perp}\left\lbrace\hat{a}^\dagger_k\hat{a}_k\hat{\mathcal{Q}}_{N-N_0}\frac{{\rm e^{-\beta\hat{\mathcal{H}}_\perp}}}
{\mathcal{Z}_\perp(N-N_0)}\hat{\mathcal{Q}}_{N-N_0}\right\rbrace
\label{occs_NC}
\end{equation} 
\end{widetext}
describes the average many particle occupation of a non-condensate single particle state $\ket{\Psi_{k}}$, given that 
$(N-N_{0})$ particles are in the non-condensate, and given a temperature $T$ of the external heat reservoir.
For the explicit derivation of analytical expressions for 
the occupation numbers $\langle N_{k}\rangle(N-N_{0},T)$, see appendix~\ref{appendix_occs}.

Anti-normally ordered products of two point correlation functions of two non-condensate fields 
in the interaction picture arising in Eqs.~(\ref{deco_corr2}) and (\ref{deco_corr1}) 
can be obtained correspondingly, turning into 
\begin{widetext}
%\vspace{-0.5cm}
\begin{equation}
 \left\langle\hat{\Psi}_{\perp}(\vec{\mbf{r}},\tau)
\hat{\Psi}^{\dagger}_{\perp}(\vec{\mbf{r}}^{\prime},0)\right\rangle^{(N-N_{0})}_{\mathcal{F}_\perp} =
\sum_{k\ne0}\Psi_{k}(\vec{\mbf{r}})\Psi^{\star}_{k}(\vec{\mbf{r}}^{\prime})
\left[\langle N_k\rangle(N-N_{0},T)+1\right]~{\rm exp}\left[\frac{i\epsilon_{k}\tau}{\hbar}\right]\ ,
\label{twopoint_sum2}
\end{equation}
where we have used that
\begin{equation} 
\left\langle\hat{a}_{k}\hat{a}_{l}^{\dagger}
\right\rangle^{(N-N_{0})}_{\mathcal{F}_\perp}
= \left[\langle N_k\rangle(N-N_{0},T)+1\right]\delta_{kl}\ .
\end{equation}
Hence, we find for normally and anti-normally ordered 
two point correlation functions with respect to single particle processes:
\begin{equation}
\begin{split}
\mathcal{G}^{(+)}_{\rightsquigarrow}(\vec{\mbf{r}},\vec{\mbf{r}}^{\prime},N-N_{0},T,\tau) &= 2\sum_{k,l,m\ne0}
\Psi_{k}^{\star}(\vec{\mbf{r}})\Psi_{k}(\vec{\mbf{r}}^{\prime})\Psi_{l}^{\star}(\vec{\mbf{r}})\Psi_{l}(\vec{\mbf{r}}^{\prime})
\Psi_{m}^{\star}(\vec{\mbf{r}})\Psi_{m}(\vec{\mbf{r}}^{\prime})\left[\langle N_k\rangle(N-N_{0},T)+1\right]\times\\
&\times \langle N_l\rangle(N-N_{0},T)
\langle N_m\rangle(N-N_{0},T)~{\rm exp}\left[\frac{i\left(\epsilon_{k}-\epsilon_l-\epsilon_m\right)\tau}{\hbar}\right] +\\
& + 4\sum_{k,l,m\ne0}
|\Psi_{k}(\vec{\mbf{r}})|^2 \Psi_{l}^{\star}(\vec{\mbf{r}})\Psi_{l}(\vec{\mbf{r}}^{\prime})
|\Psi_{m}(\vec{\mbf{r}}^\prime)|^2\langle N_k\rangle(N-N_{0},T)\times\\
&\times \langle N_l\rangle(N-N_{0},T)
\langle N_m\rangle(N-N_{0},T)~{\rm exp}\left[\frac{-i\epsilon_l\tau}{\hbar}\right]\ .
 \end{split}
\label{gpoint1}
\end{equation}
\begin{equation}
\begin{split}
\mathcal{G}^{(-)}_{\rightsquigarrow}(\vec{\mbf{r}},\vec{\mbf{r}}^{\prime}&,N-N_{0},T,\tau) = 2\sum_{k,l,m\ne0}
\Psi_{k}^{\star}(\vec{\mbf{r}})\Psi_{k}(\vec{\mbf{r}}^{\prime})\Psi_{l}^{\star}(\vec{\mbf{r}})\Psi_{l}(\vec{\mbf{r}}^{\prime})
\Psi_{m}^{\star}(\vec{\mbf{r}})\Psi_{m}(\vec{\mbf{r}}^{\prime})\langle N_k\rangle(N-N_{0},T)\times\\
&\times \left[\langle N_l\rangle(N-N_{0},T)+1\right]
\left[\langle N_m\rangle(N-N_{0},T)+1\right]~{\rm exp}\left[\frac{-i\left(\epsilon_{k}-\epsilon_l-\epsilon_m\right)\tau}{\hbar}\right]+ \\
& +4\sum_{k,l,m\ne0}
|\Psi_{k}(\vec{\mbf{r}})|^2\Psi_{l}^{\star}(\vec{\mbf{r}})\Psi_{l}(\vec{\mbf{r}}^{\prime})
|\Psi_{m}(\vec{\mbf{r}})|^2\langle N_k\rangle(N-N_{0},T)\times\\
&\times \left[\langle N_l\rangle(N-N_{0},T)+1\right]
\langle N_m\rangle(N-N_{0},T)~{\rm exp}\left[\frac{i \epsilon_l \tau}{\hbar}\right]\ .
\end{split}
\label{gpoint2}
\end{equation}
\end{widetext}
Integration of $\mathcal{G}^{(\pm)}_{\rightsquigarrow}(\vec{\mbf{r}},\vec{\mbf{r}}^{\prime},N-N_{0},T,\tau)$ over the time interval 
$\tau$, multiplied by $\Psi_0(\vec{\mbf{r}})\Psi^\star_0(\vec{\mbf{r}}^{\prime}){\rm exp}[\pm i\mu_0\tau]$, which arises from the backswitch 
of the condensate fields from the interaction picture, leads to the 
single particle loss and feeding rates in Eq.~(\ref{rate1}).

Next, we decompose the correlation functions 
for pair events, $\mathcal{G}^{(\pm)}_{\leftrightsquigarrow}(\vec{\mbf{r}},\vec{\mbf{r}}^{\prime},N-N_{0},T,\tau)$.
%i.e. one two point correlation function 
%of a product of four non-condensate fields into a product of two time ordered 
%two point correlation functions of two non-condensate fields.
Using Eq.~(\ref{twopoint_sum},~\ref{twopoint_sum2}), the normally ordered correlation function for pair events is given by:
\begin{widetext}
%\vspace{-0.4cm}
\begin{equation}
\begin{split}
&\mathcal{G}^{(+)}_{\leftrightsquigarrow}(\vec{\mbf{r}},\vec{\mbf{r}}^{\prime},N-N_{0},T,\tau) = \left\langle\hat{\Psi}_{\perp}^{\dagger}(\vec{\mbf{r}},\tau)
\hat{\Psi}_{\perp}^{\dagger}(\vec{\mbf{r}},\tau)
\hat{\Psi}_{\perp}(\vec{\mbf{r}}^{\prime},0)\hat{\Psi}_{\perp}(\vec{\mbf{r}}^{\prime},0)
\right\rangle^{(N-N_{0})}_{\mathcal{F}_\perp} = \\
&2\left\langle\hat{\Psi}_{\perp}^{\dagger}(\vec{\mbf{r}},\tau)
\hat{\Psi}_{\perp}(\vec{\mbf{r}}^{\prime},0)\right\rangle^{(N-N_{0})}_{\mathcal{F}_\perp}
\left\langle\hat{\Psi}_{\perp}^{\dagger}(\vec{\mbf{r}},\tau)
\hat{\Psi}_{\perp}(\vec{\mbf{r}}^{\prime},0)\right\rangle^{(N-N_{0})}_{\mathcal{F}_\perp}=\\
&2\sum_{k,l\ne0}
\Psi_{k}^{\star}(\vec{\mbf{r}})\Psi_{k}(\vec{\mbf{r}}^{\prime})\Psi_{l}^{\star}(\vec{\mbf{r}})\Psi_{l}(\vec{\mbf{r}}^{\prime})
\langle N_k\rangle(N-N_{0},T)\langle N_l\rangle(N-N_{0},T)~{\rm exp}\left[\frac{+i\left(\epsilon_{k}+\epsilon_l\right)\tau}{\hbar}\right]\ . 
\end{split}
\label{deco_pair_corr1}
\end{equation}
The anti-normally ordered pair correlation function 
$\mathcal{G}^{(-)}_{\leftrightsquigarrow}(\vec{\mbf{r}},\vec{\mbf{r}}^{\prime},N-N_{0},T,\tau)$ can be 
decomposed similarly: 
\begin{equation}
\begin{split}
&\mathcal{G}^{(-)}_{\leftrightsquigarrow}(\vec{\mbf{r}},\vec{\mbf{r}}^{\prime},N-N_{0},T,\tau) = \left\langle\hat{\Psi}_{\perp}(\vec{\mbf{r}},\tau)
\hat{\Psi}_{\perp}(\vec{\mbf{r}},\tau)
\hat{\Psi}_{\perp}^{\dagger}(\vec{\mbf{r}}^{\prime},0)
\hat{\Psi}^{\dagger}_{\perp}(\vec{\mbf{r}}^{\prime},0)
\right\rangle^{(N-N_{0})}_{\mathcal{F}_\perp} = \\
&2\left\langle\hat{\Psi}_{\perp}(\vec{\mbf{r}},\tau)
\hat{\Psi}_{\perp}^{\dagger}(\vec{\mbf{r}}^{\prime},0)\right\rangle^{(N-N_{0})}_{\mathcal{F}_\perp}
\left\langle\hat{\Psi}_{\perp}(\vec{\mbf{r}},\tau)
\hat{\Psi}^{\dagger}_{\perp}(\vec{\mbf{r}}^{\prime},0)\right\rangle^{(N-N_{0})}_{\mathcal{F}_\perp}=\\
&2\sum_{k,l\ne0}
\Psi_{k}^{\star}(\vec{\mbf{r}})\Psi_{k}(\vec{\mbf{r}}^{\prime})\Psi_{l}^{\star}(\vec{\mbf{r}})\Psi_{l}(\vec{\mbf{r}}^{\prime})
\left[\langle N_k\rangle(N-N_{0},T)+1\right]\left[\langle N_l\rangle(N-N_{0},T)+1\right]~{\rm exp}\left[\frac{-i\left(\epsilon_{k}+\epsilon_l\right)\tau}{\hbar}\right]\ , 
\end{split}
\label{deco_pair_corr2}
\end{equation}
which again, after multiplication with $\Psi^2_0(\vec{\mbf{r}})\left(\Psi^\star_0(\vec{\mbf{r}}^{\prime})\right)^2{\rm exp}[\pm 2i\mu_0\tau]$ 
and integration over $\tau$, turns into the pair feeding and loss rates in Eq.~(\ref{rate2}).
\end{widetext}

\section{Single particle non-condensate occupations}
\label{appendix_occs}
The state of the non-condensate in Eq.~(\ref{Bose_state_final}) allows to determine the 
average number of particles, $\langle N_k \rangle = 
\langle N_k \rangle(N-N_0,T)$ in Eq.~(\ref{occs_NC}), for each particular 
non-condensate single particle mode 
$\ket{\Psi_{k}}$, given that 
$N_0$ particles populate the condensate mode, 
and consequently $(N-N_0)$ particles populate the non-condensate single particle modes.  
According to Eq.~(\ref{occs_NC}), we hence consider the expectation value of the number operator $\hat{N}_k$ 
in a non-condensate state of $(N-N_0)$ particles, which leads to 
\begin{widetext}
%\vspace{-1.2cm}
\begin{equation}
 \langle N_k \rangle(N-N_0,T) 
= \mathcal{Z}^{-1}_\perp(N-N_0)\sum_{\lbrace N_k\rbrace}^{(N-N_0)}
N_{k}{\rm exp}\left[-\beta\sum_{k\ne0}\epsilon N_k\right]\ , 
\label{expectation_NC}
\end{equation}
\end{widetext}
where $\mathcal{Z}_\perp(N-N_0)$ is the partition function of $(N-N_0)$ indistinguishable 
particles in the non-condensate in Eq.~(\ref{partition}).
In terms of the partial partition sum, $\mathcal{Z}_\perp^{(k)}(N-N_0)$~\cite{Statistics}, 
which excludes the sum over one particular non-condensate single particle mode $\ket{\Psi_k}$, 
Eq.~(\ref{expectation_NC}) can be written as 
\begin{widetext}
%\vspace{-0.8cm}
\begin{equation}
 \langle N_k \rangle(N-N_0,T) = 
\mathcal{Z}_\perp^{-1}(N-N_0)\sum_{N_{k}=0}^{(N-N_0)}
N_{k}~{\rm exp}\left[-\beta\epsilon_{k}N_{k}\right]\mathcal{Z}_\perp^{(k)}(N-N_0-N_{k})\ .
\label{expectation_NC2}
\end{equation}
For small enough $N_{k}$ (it suffices to start at $N_{k}=1$ and to
determine $\mathcal{Z}_\perp^{(k)}(N-N_0-N_{k})$ 
iteratively), we can expand
\begin{equation}
{\rm ln}\left[\mathcal{Z}_\perp^{(k)}(N-N_0-1)\right]\simeq{\rm ln}\left[\mathcal{Z}_\perp^{(k)}(N-N_0)\right] - 
\alpha^{(k)}_{\perp}(N-N_0,T)\ ,
\label{iteration_pre}
\end{equation}
which introduces the parameter 
\begin{equation}
 \alpha^{(k)}_{\perp}(N-N_0,T) = \frac{\partial {\rm ln}\left[\mathcal{Z}_\perp^{(k)}(N-N_0)\right]}{\partial(N-N_0)}\ .
\label{alpha}
\end{equation}
From Eq.~(\ref{iteration_pre}), we hence find the recursion relation
\begin{equation}
\frac{\mathcal{Z}_\perp^{(k)}(N-N_0-1)}{\mathcal{Z}_\perp^{(k)}(N-N_0)}=
{\rm exp}\left[-\alpha^{(k)}_{\perp}(N-N_0,T)\right]\ ,
\label{iteration}
\end{equation}
between the partial partition sums 
$\mathcal{Z}_\perp^{(k)}(N-N_0)$ of $(N-N_0)$ and $\mathcal{Z}_\perp^{(k)}(N-N_0-1)$ of 
$(N-N_0-1)$ non-condensate particles. 
Multiple iteration of Eq.~(\ref{iteration}) hence leads to 
\begin{equation}
\frac{\mathcal{Z}_\perp^{(k)}(N-N_0-N_k)}{\mathcal{Z}_\perp^{(k)}(N-N_0)} = {\rm exp}
\left[-N_k\alpha^{{\rm (k)}}_{\perp}(N-N_0,T)\right]\ ,
\label{iteration2}
\end{equation}
and Eq.~(\ref{expectation_NC2}) turns into 
\begin{equation}
 \langle N_k \rangle(N-N_0,T) 
= \frac{\mathcal{Z}_\perp^{(k)}(N-N_0)}{\mathcal{Z}_\perp(N-N_0)}\sum_{N_{k}=0}^{(N-N_0)}
N_{k}{\rm exp}\left[-\left(\beta\epsilon_{k}+\alpha^{{\rm (k)}}_{\perp}(N-N_0,T)\right)N_{k}\right]\ .
\label{expectation_NC3}
\end{equation}
It remains to apply the same procedure 
to the partition function $\mathcal{Z}_\perp(N-N_0)$. 
Using the decomposition in Eq.~(\ref{expectation_NC3}) 
and applying Eq.~(\ref{iteration2}), one finds that 
\begin{equation}
\mathcal{Z}_\perp(N-N_0) = \mathcal{Z}_\perp^{(k)}(N-N_0)\sum_{N_{k}=0}^{(N-N_0)}
{\rm exp}\left[-\left(\beta\epsilon_{k}+\alpha^{{\rm (k)}}_{\perp}(N-N_0,T)\right)N_{k}\right]\ .
\label{partition1}
\end{equation}
Setting Eq.~(\ref{partition1}) into Eq.~(\ref{expectation_NC}), 
the expectation value of particle number occupations of a particular non-condensate single particle 
state $\ket{\Psi_k}$, given that $(N-N_0)$ particles are in the non-condensate, 
is given by  
\begin{equation}
\langle N_k \rangle(N-N_0,T) 
= \frac{1}{{\rm exp}\left[\beta\epsilon_{k}+\alpha^{{\rm (k)}}_\perp(N-N_0,T)\right]-1}\ .
\label{eq1}
\end{equation}
\end{widetext}
We now use that the parameter $\alpha^{(k)}_{\perp}(N-N_0,T)$ is approximately 
independent of the state $k$~\cite{Statistics}, i.e. the change of non-condensate single particle 
number occupations 
during condensation is described by one 
single parameter, $\alpha^{{\rm (k)}}\simeq\alpha_\perp(N-N_0,T)$, which is determined by 
the constraint of particle number conservation, as spelled out by the implicit equation
\begin{widetext}
%\vspace{-0.3cm}
\begin{equation}
\sum_{k\ne0}\langle N_k \rangle(N-N_0,T) = 
\sum_{k\ne0}\frac{1}{{\rm exp}[\beta\epsilon_{k}+\alpha_\perp(N-N_0,T)]-1} = (N-N_0)\ . 
\label{chemical}
\end{equation}
\end{widetext}
As evident from Eq.~(\ref{alpha}), and the fact that each subspace of $(N-N_0)$ particles is in a thermal state, the 
parameter $\alpha_\perp(N-N_0,T)$ can be interpreted as the ratio of the 
non-condensate chemical potential for a state of $(N-N_0)$ atoms, to the 
thermal energy $\beta^{-1}$. 
Hence, from the definition in Eq.~(\ref{alpha}), we see that $\alpha_\perp(N-N_0,T)$ 
is, upon a constant, nothing more than the derivative of the Helmholtz free energy, 
$\alpha_\perp(N-N_0,T)=-\beta^{-1}{\rm ln}\mathcal{Z}_\perp(N-N_0)$
of the $(N-N_0)$ particles in the non-condensate~\cite{Statistics}, related 
to $\mu_\perp(N-N_0,T)$ by
\begin{equation}
 \alpha_\perp(N-N_0,T)=-\beta\mu_\perp(N-N_0,T)\ ,
 \label{NC_ch}
\end{equation}
which introduces the non-condensate chemical potential $\mu_\perp(N-N_0,T)$.

\end{document}